\documentclass[prx,letterpaper,nobalancelastpage,twocolumn,groupedaddress,nofootinbib,longbibliography]{revtex4-2}

\usepackage{xr-hyper}
\usepackage{graphicx}
\usepackage{amsmath}
\DeclareMathOperator{\sinc}{sinc}
\usepackage{bbold}
\usepackage{amssymb}
\usepackage[english]{babel}
\usepackage{color}
\usepackage[version=4]{mhchem}
\usepackage[hidelinks]{hyperref}
\usepackage{dsfont}
\usepackage{placeins}
\usepackage{mathtools}
\usepackage[normalem]{ulem}
\usepackage{varwidth}
\usepackage{siunitx}
\usepackage{gensymb}
\usepackage{bibunits}
\usepackage{upgreek}
\setcitestyle{super}

\usepackage{soul}
\newcommand{\ket}[1]{| #1 \rangle}

\newcommand{\Caltech}{California Institute of Technology, Pasadena, CA 91125, USA}

\usepackage{caption}

\DeclareCaptionLabelSeparator{bar}{ \textbf{\textbar}~}
\usepackage{enumitem}
\setlist{nolistsep}

\let\oldchi\chi
\renewcommand{\chi}{
  \raisebox{0.44ex}{$\oldchi$}
}

\makeatletter
\newcommand*{\addFileDependency}[1]{
\typeout{(#1)}
\@addtofilelist{#1}
\IfFileExists{#1}{}{\typeout{No file #1.}}
}\makeatother

\defaultbibliographystyle{modified-ref.bst}
\defaultbibliography{references.bib,additional-ref.bib}

\begin{document}

\begin{bibunit}

\title{A tweezer array with 6100 highly coherent atomic qubits}

\author{Hannah J. Manetsch}\thanks{These authors contributed equally to this work}
\author{Gyohei Nomura}\thanks{These authors contributed equally to this work}
\author{Elie Bataille}\thanks{These authors contributed equally to this work}
\author{Kon H. Leung}
\author{Xudong Lv}
\author{Manuel Endres}\email{mendres@caltech.edu}
\affiliation{\Caltech}

\begin{abstract}
Optical tweezer arrays~\cite{browaeys_many-body_2020,kaufman_quantum_2021} have transformed atomic and molecular physics, now forming the backbone for a range of leading experiments in quantum computing~\cite{saffman_quantum_2016,henriet_quantum_2020, bluvstein_quantum_2022,graham_multi-qubit_2022,ma_universal_2022,finkelstein_universal_2024}, simulation~\cite{bernien_probing_2017,browaeys_many-body_2020,ebadi_quantum_2021,scholl_quantum_2021,shaw_benchmarking_2024}, and metrology~\cite{norcia_seconds-scale_2019,madjarov_atomic-array_2019,young_half-minute-scale_2020}. Typical experiments trap tens to hundreds of atomic qubits~\cite{young_half-minute-scale_2020,scholl_quantum_2021,huft_simple_2022,singh_mid-circuit_2023,tao_high-fidelity_2024,bluvstein_logical_2024,shaw_benchmarking_2024}, and recently systems with around one thousand atoms were realized without defining qubits or demonstrating coherent control~\cite{pause_supercharged_2024,norcia_iterative_2024_mod,gyger_continuous_2024}. However, scaling to thousands of atomic qubits with long coherence times, low-loss, and high-fidelity imaging is an outstanding challenge and critical for progress in quantum science, particularly towards quantum error correction~\cite{preskill_quantum_2018,xu_constant-overhead_2024}. Here, we experimentally realize an array of optical tweezers trapping over 6,100 neutral atoms in around 12,000 sites, simultaneously surpassing state-of-the-art performance for several metrics that underpin the success of the platform. Specifically, while scaling to such a large number of atoms, we demonstrate a coherence time of 12.6(1) seconds, a record for hyperfine qubits in an optical tweezer array~\cite{bluvstein_quantum_2022,graham_multi-qubit_2022}. We show room-temperature trapping lifetimes of ${\sim}$23 minutes, enabling record-high imaging survival~\cite{covey_2000-times_2019,blodgett_imaging_parsed_2023} of $99.98952(1)\%$ with an imaging fidelity of over $99.99\%$. We present a plan for zone-based quantum computing~\cite{bluvstein_quantum_2022,bluvstein_logical_2024} and demonstrate necessary coherence-preserving qubit transport and pick-up/drop-off operations on large spatial scales, characterized through interleaved randomized benchmarking.
Our results, along with recent developments~\cite{evered_high-fidelity_2023,ma_high-fidelity_2023, zhang_scaled_2024, finkelstein_universal_2024}, indicate that universal quantum computing~\cite{bluvstein_quantum_2022,graham_multi-qubit_2022,ma_universal_2022,finkelstein_universal_2024} and quantum error correction~\cite{preskill_quantum_2018,xu_constant-overhead_2024} with thousands to tens of thousands of physical qubits could be a near-term prospect. 
\end{abstract}

\maketitle

Optical tweezer arrays~\cite{browaeys_many-body_2020,kaufman_quantum_2021} have transformed atomic and molecular physics experiments by simplifying detection and enabling individual-particle control~\cite{barredo_atom-by-atom_2016,endres_atom-by-atom_2016, kim_situ_2016,liu_building_2018,anderegg_optical_2019}, resulting in rapid, recent progress in quantum computing~\cite{saffman_quantum_2016,henriet_quantum_2020, bluvstein_quantum_2022,graham_multi-qubit_2022,ma_universal_2022,finkelstein_universal_2024}, simulation~\cite{bernien_probing_2017,browaeys_many-body_2020,ebadi_quantum_2021,scholl_quantum_2021,shaw_benchmarking_2024}, and metrology~\cite{norcia_seconds-scale_2019,madjarov_atomic-array_2019,young_half-minute-scale_2020}. In this context, each atom typically encodes a single qubit that is controlled with electromagnetic fields, and ideally features long coherence times to enable these applications with high fidelity. Such optically trapped atomic qubits, including lattices~\cite{miroshnychenko_atom-sorting_2006,weitenberg_single-spin_2011} and lattice-tweezer hybrid systems~\cite{young_half-minute-scale_2020,tao_high-fidelity_2024,gyger_continuous_2024}, coexist with other platforms that have single-qubit control and readout, including ion traps~\cite{bruzewicz_trapped-ion_2019} and superconducting qubits~\cite{kjaergaard_superconducting_2020}.

There are important incentives to scale up such fully programmable qubit platforms. Optical atomic clocks gain stability with increasing atom number~\cite{ludlow_optical_2015,rosenband_exponential_2013}, while quantum simulation experiments benefit from thousands of qubits to explore emergent collective phenomena~\cite{orourke_entanglement_2023,julia-farre_amorphous_2024} or demonstrate verifiable quantum advantage~\cite{haferkamp_closing_2020,ringbauer_verifiable_2025}. Most critically, quantum error correction (QEC) demands both large system sizes and exceptional fidelities: even the most resource-efficient protocols require several thousand physical qubits operating with error rates below $10^{-3}$ to encode $>$100 logical qubits~\cite{bravyi_high-threshold_2024,xu_constant-overhead_2024}. This represents a fundamental scalability challenge that has limited the practical impact of quantum technologies.

Current universal quantum computing architectures, such as those based on neutral atoms~\cite{bluvstein_quantum_2022,graham_multi-qubit_2022,ma_universal_2022,finkelstein_universal_2024}, ions~\cite{moses_race_2023}, and superconducting qubits~\cite{kim_evidence_2023}, typically operate with tens to hundreds of qubits. While most platforms suffer from increasingly deleterious effects as system size grows~\cite{kjaergaard_superconducting_2020,bruzewicz_trapped-ion_2019}, neutral atoms in optical tweezer arrays offer a promising solution for rapid scalability in the near term thanks to a programmable architecture adaptable to larger system sizes.

Universal quantum computing capabilities with neutral atoms have recently been realized in optical tweezer array systems, based on demonstrations of individual qubit addressing~\cite{huie_repetitive_2023,lis_midcircuit_2023,norcia_midcircuit_2023,shaw_multi-ensemble_2024}, high-fidelity entangling gates~\cite{evered_high-fidelity_2023,finkelstein_universal_2024}, coherence-preserving dynamical reconfigurability~\cite{beugnon_two-dimensional_2007,bluvstein_quantum_2022,finkelstein_universal_2024}, and ancilla-based mid-circuit measurement~\cite{singh_mid-circuit_2023,bluvstein_logical_2024,finkelstein_universal_2024}. Very recently, tweezer systems with about a thousand atoms have been realized in a discontiguous array based on interleaved microlens elements~\cite{pause_supercharged_2024}, and via repeated reloading from a reservoir~\cite{norcia_iterative_2024_mod,gyger_continuous_2024}; none of these experiments, however, report control of qubits, measurement of coherence times, or coherence-preserving transport.

\begin{figure*}[t!]
\centering
    \includegraphics[width=\textwidth]{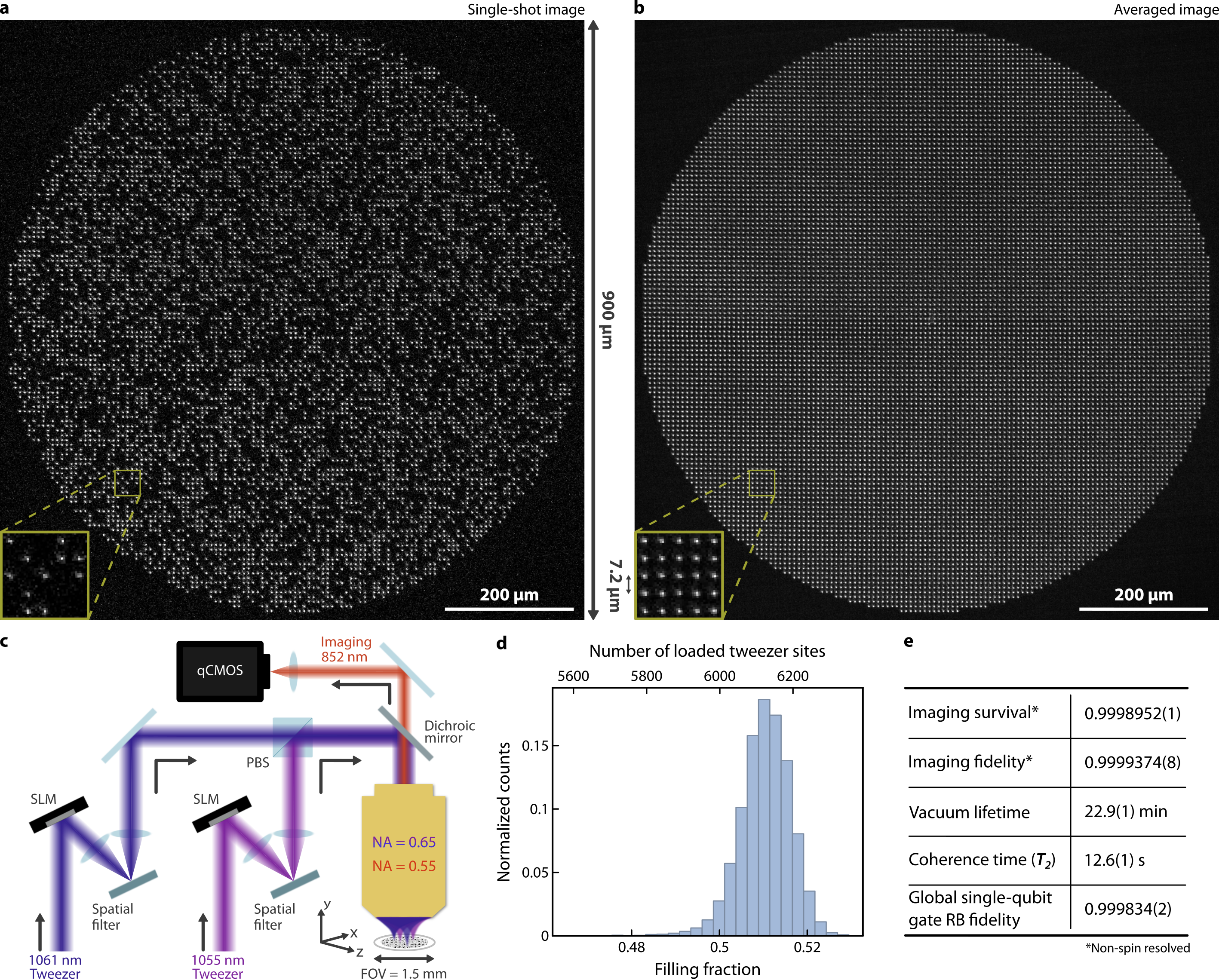}
        \caption{\textbf{Large-scale tweezer array.} \textbf{a,} Representative single-shot image of single cesium atoms across a 11,998-site tweezer array. Inset: magnified view of a subsection of the stochastically loaded array. \textbf{b,} Averaged image of single atoms across a 11,998-site tweezer array. Inset: magnified view of a subsection of the averaged array. Atoms are spaced by $7.2 \ \upmu$m and held in 1061 nm and 1055 nm optical tweezers. The contrast is enhanced for visual clarity. \textbf{c,} Schematic of the optical tweezer array generation. Tweezer arrays, generated by two spatial light modulators (SLM), at 1061 nm and 1055 nm are combined with orthogonal polarization, and focused through an objective with a numerical aperture (NA) of 0.65 and a field of view (FOV) of 1.5 mm in diameter. The direction of gravity is along $\hat{y}$. We collect scattered photons from single atoms through the same objective and image them on a qCMOS camera. \textbf{d,} Histogram of filling fraction. We load 6,139 single atoms on average per experimental iteration (51.2\% of the array on average), with a relative standard deviation of 1.13\% over 16,000 iterations. \textbf{e,} Summary of the key metrics demonstrated in this work.}

    \vspace{-0.5cm}
    \label{Fig1}
\end{figure*}

Here, we demonstrate a tweezer array with 11,998 sites, that traps over 6,100 atomic qubits, simultaneously matching or surpassing state-of-the-art values for metrics underpinning the usefulness of the platform, including hyperfine qubit coherence time, trapping lifetime in a room-temperature apparatus, coherent transport distance and fidelity, trap transfer fidelity, as well as combined imaging fidelity and survival (Fig.~\ref{Fig1}). Our results have implications for the aforementioned applications in quantum science, in particular, concerning large-scale quantum computing and error correction, as discussed in more detail below.

\vspace{0.25cm}
\noindent\textbf{Summary of approach and results}\\
Our approach leverages high-power trapping of single cesium-133 atoms at far-off-resonant wavelengths in a specially designed, room-temperature vacuum chamber (Methods, Ext. Data Fig.~\ref{ext_fig1}a), enabling low-loss, high-fidelity imaging in combination with long hyperfine coherence times at the scale of 6,100 qubits (Fig.~\ref{Fig1}e). Specifically, we demonstrate single-atom imaging with a survival probability of $99.98952(1)\%$ and a fidelity of $99.99374(8)\%$, surpassing the state-of-the-art achieved in much smaller arrays\cite{covey_2000-times_2019}. This, alongside a 22.9(1)-minute vacuum-limited lifetime in our room-temperature apparatus~\cite{schymik_single_2021}---significantly longer than typical state-of-the-art vacuum lifetimes for tweezer arrays in room-temperature apparatuses---provides realistic timescales for array operations in large scale arrays with minimal loss; e.g., for atomic rearrangement~\cite{barredo_atom-by-atom_2016,endres_atom-by-atom_2016, kim_situ_2016}.

Importantly, we further demonstrate a coherence time of 12.6(1) s, a record for a hyperfine qubit tweezer array, surpassing previous values by almost an order of magnitude~\cite{bluvstein_quantum_2022,graham_multi-qubit_2022} and approaching results for a single hyperfine qubit in a customized blue-detuned trap~\cite{tian_extending_2024}, alkali atoms in an optical lattice~\cite{wu_sterngerlach_2019}, and nuclear qubits in a tweezer array~\cite{barnes_assembly_2022}. We also show a single-qubit gate fidelity of $99.9834(2)\%$ measured with global randomized benchmarking.

Finally, we demonstrate coherent atomic transport across 610 $\upmu$m with a fidelity of $\sim$99.95\% and coherent transfer between static and dynamic traps with a fidelity of 99.81(3)\%. These together form crucial ingredients for scaling atomic quantum processors in a zone-based architecture, with a detailed plan laid out further in the Supplementary Information (SI). Our results indicate that quantum computing with 6,000 atomic qubits is a near-term prospect, providing a path towards QEC with hundreds of logical qubits~\cite{xu_constant-overhead_2024}.

\vspace{0.25cm}
\noindent\textbf{Large-scale optical tweezer generation}\\
To scale the optical tweezer array platform, while extending hyperfine coherence times, we generate traps using near-infrared wavelengths, far-detuned from dominant electric-dipole transitions, thus minimizing photon scattering and dephasing processes~\cite{ozeri_hyperfine_2005,ozeri_errors_2007}. Cesium atoms possess the highest polarizability among the stable alkali-metal atoms at near-infrared wavelengths where commercial fiber amplifiers provide continuous-wave laser powers that exceed 100 W. Thus, a large number of traps can be created with sufficient depth. A representative single shot image of the array is shown in Fig.~\ref{Fig1}a, and an averaged image is shown in Fig.~\ref{Fig1}b.

The atoms are spaced by ${\sim}7.2\,\mathrm{\upmu m}$ and held in traps at 1055 nm and 1061 nm, generated using spatial light modulators (SLMs), whose hologram phases are optimized with a weighted Gerchberg-Saxton algorithm~\cite{kim_gerchberg-saxton_2019, kim_large-scale_2019} to uniformize the tweezer trap depth (Methods). The tweezer light is combined with polarization and focused through a high numerical aperture objective with a large field of view of 1.5 mm in diameter, usable for atom trapping and manipulation (Fig.~\ref{Fig1}c).

The tweezers are created with 130 W of optical power generated from fiber amplifiers. After transmission through the optical path, around 35-40 W reaches the objective, and from trap parameter measurements (see ``Tweezer generation" Methods section), we estimate ${\sim}1.4$ mW to be used per tweezer at the atom plane. We measure an average trap depth of $k_B\times 0.18(2)$~mK, with a standard deviation of $11.4\%$ across all sites (Ext. Data Fig.~\ref{fig:loading-ed}d), enabling consistent loading probability per site.

\vspace{0.25cm}
\noindent\textbf{Loading and imaging single atoms}\\
\begin{figure}[t!]
    \centering
    \includegraphics[width=85mm]{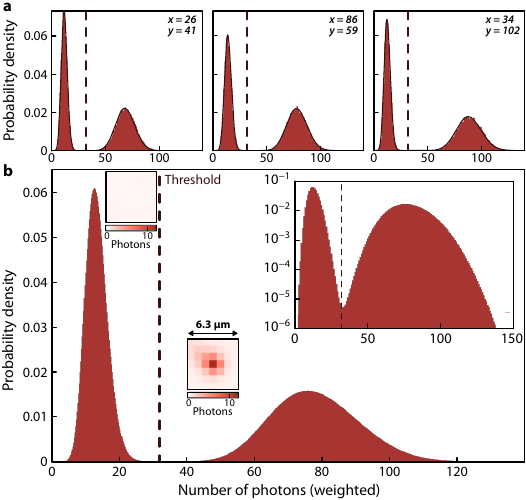}
    \caption{\textbf{High-fidelity atom detection in a large-scale tweezer array.} Imaging histogram showing the number of photons collected per site and per image. Note that the horizontal axes are weighted photon counts (see text); for non-weighted photon counts, see Ext. Data Fig.~\ref{fig:imaging-ed}b.
    \textbf{a}, Imaging histogram of three randomly selected sites in the array (where $x$ and $y$ respectively denote the horizontal and vertical site indices in the array), and \textbf{b}, averaged over all sites in the array. Per-site histograms are fitted with a Poissonian model that integrates losses during imaging (Methods). The wide separation of peaks for empty and filled tweezers enables the high imaging fidelity presented in this work. The binarization threshold used to determine tweezer occupation is indicated by the vertical dashed line, and the average point-spread functions for the two classifications (atom absent and atom detected) are shown next to their corresponding peaks. Note that we detect no more than one atom in each tweezer. Inset: the same histogram presented with a log-scale vertical axis.  The weighted average relative error bar per bin is 0.08\% (0.05\% for the log-scale inset due to the smaller number of bins).} 
    \vspace{-0.5cm}
    \label{Fig2}
\end{figure}

We demonstrate uniform loading and high imaging fidelity across the sites in the array. To load single atoms in the tweezers, we cool and then parity-project~\cite{schlosser_sub-poissonian_2001} from a ${\sim}1.6$~mm $1/e^2$ diameter magneto-optical trap.
Before imaging the atoms, we use a multi-pronged approach to filter out atoms in spurious off-plane traps, residual from the SLM tweezer creation (Methods).

We then zero the magnetic field and apply two-dimensional polarization gradient cooling (2D PGC) in the atom array plane ($x$-$z$ plane in Fig.~\ref{Fig1}c) for fluorescence imaging of single atoms, which simultaneously cools the atoms. Imaging light is applied for 80~ms, and photons are imaged on a quantitative CMOS camera (qCMOS). We find that each site has a loading probability of 51.2\% with a relative standard deviation of 3.4\% across the sites, demonstrating uniform filling of single atoms (Ext. Data Fig.~\ref{fig:loading-ed}c). This allows us to load over 6,100 sites on average in each iteration (Fig.~\ref{Fig1}d).

We distinguish atomic presence in the array with high fidelity. Each image undergoes a binarization procedure (Methods) whereby each site is attributed a value of 0 (no atom detected) or 1 (one atom detected). We weight the collected photons in a $7\times 7$-pixel box centered around each site~\cite{cooper_alkaline-earth_2018}, so as to add more weight to pixels close to the center of each site's point-spread function (Ext. Data Fig.~\ref{fig:imaging-ed}a). The resulting signal is compared with a threshold to determine if an atom is present or not (Fig.~\ref{Fig2}). 

We characterize the imaging fidelity, defined as the probability of correctly labeling atomic presence in a site, with a model-free approach, where no assumption is made about the photon distribution from Fig.~\ref{Fig2}. To this end, we identify anomalous series of binary outputs~\cite{norcia_microscopic_2018} in three consecutive images. For instance, $0\rightarrow1\rightarrow1$ would point to a false negative event in the first image, while $1\rightarrow1\rightarrow0$ could be due to atom loss during the second image or a false negative event in the third one. This approach allows us to precisely decouple inherent atom loss from false negatives or positives. From this we find an imaging fidelity of 99.99374(8)\% (note that we excise the first image, which we find has slightly lower fidelity and survival probability; Methods). Crucial to this result is the homogeneous photon scattering rate across the array (Ext. Data Fig.~\ref{fig:imaging-ed}d) and the consistency of the point-spread function across the array (waist radius of 1.7 pixels with a standard deviation of 0.2 pixels). Consistent imaging parameters across the array further are evidenced in that we find that treating each site with an individual threshold only marginally improves the imaging fidelity to 99.9939(1)\%. Finally, we estimate that the imaging fidelity in the absence of atomic loss would be closer to 99.999\% (Methods).

\begin{figure}[t!]
    \centering
    \includegraphics[width=89mm]{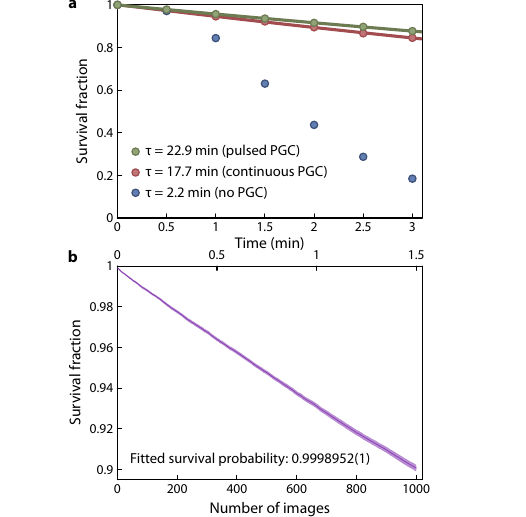}
        \caption{\textbf{Long vacuum-limited lifetime and high imaging survival probability} \textbf{a,} Vacuum-limited lifetime. Array-averaged survival fraction as a function of hold time is plotted. Three experiments are shown in the figure: pulsed cooling, continuous cooling, and no cooling. The green markers show data with a 10-ms 2D PGC block applied every 2 s during the wait time (pulsed PGC). The red markers show data with 2D PGC block continuously applied during the wait time (continuous PGC), while the blue markers show the data without cooling during the wait time (no PGC). The error bars are smaller than the markers. We find a $1/e$ lifetime of around 2.2 min without cooling. When the pulsed PGC block is applied, by fitting the data with $p(t) \propto \exp(-t/\tau)$, we find a vacuum lifetime of $\tau=22.9(1)$ min. When the 2D PGC is applied continuously, we obtain $\tau=17.7(2)$ min. \textbf{b,} Array-averaged survival fraction after many successive images. Between each image, we hold the atoms for 10 ms, without applying any cooling beams. We fit the data with $p(N) \propto p_{1}^N$, where $p(N)$ is the survival fraction after imaging $N$ times. From the fit, we find a steady-state imaging survival probability of $p_{1}=0.9998952(1)$. The light purple fill shows the estimated 68\% confidence interval.}
    \vspace{-0.5cm}
    \label{Fig3}
\end{figure}
\vfill\null

\noindent\textbf{Imaging survival and vacuum-limited lifetime}\\
The probability of losing no atom in a tweezer array during imaging and due to finite vacuum lifetime both decrease exponentially in the number of atoms in an array, making these crucial metrics to optimize for large-scale array operation. The vacuum-limited lifetime, in particular, sets an upper bound on the duration during which operations can be executed without loss of an atom in a given experimental run. This can, for example, be applied as an upper limit on the fidelity with which one can achieve a defect-free array via atom rearrangement~\cite{barredo_atom-by-atom_2016,endres_atom-by-atom_2016}. 

We probe the vacuum-limited lifetime using an empirically optimized cooling sequence consisting of a 10-ms 2D PGC cooling block every 2~s. By fitting the exponential decay of the atom survival, we find a $1/e$ lifetime of 22.9(1) min (Fig.~\ref{Fig3}a). This is a significantly long timescale compared with state-of-the-art room-temperature atomic experiments, and within a factor of five of the longest reported lifetime in a cryogenic apparatus~\cite{schymik_single_2021}. The result indicates that the probability of losing a single atom across the entire array remains under 50\% during 100 ms, a relevant timescale for dynamical array reconfiguration and quantum processor operation.

\begin{figure*}[ht!]
    \centering
    \includegraphics[width=113mm]{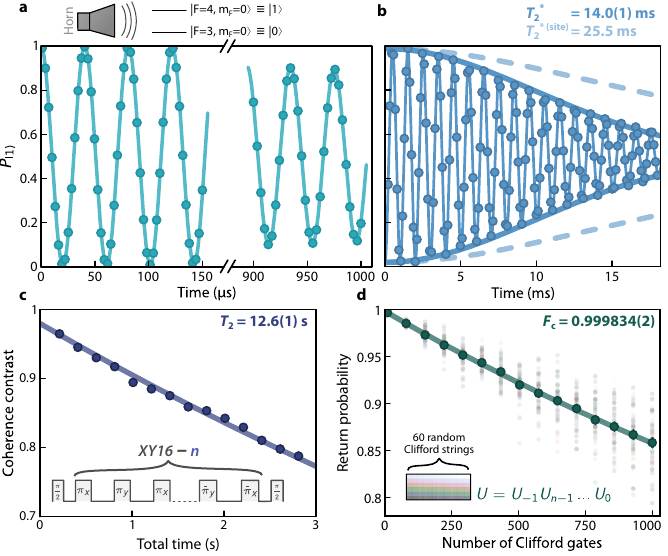}
    \caption{\textbf{Long coherence times and high-fidelity single-qubit gates in a large atom array.} \textbf{a,} Array-averaged Rabi oscillations between the hyperfine clock states $\ket{0}$ and $\ket{1}$. The fitted Rabi frequency is 24.611(1) kHz. The observed decay after several hundred microseconds stems from the spatially-varying Rabi frequency (Ext. Data Fig.~\ref{fig:uw-ed}b). \textbf{b,} Array-averaged Ramsey oscillations. During free evolution, the microwave drive field is detuned by 1 kHz, resulting in Ramsey oscillations. The characteristic decay time of these oscillations is $T_2^* = 14.0(1)\,\mathrm{ms}$ from fitting the average signal of all atoms. The light blue dashed line shows the decay time $T_2^{*(\mathrm{site})} = 25.5 \ \mathrm{ms}$ from fitting individual sites first and averaging the decay time afterwards. \textbf{c,} Measurement of the dephasing time $T_2$ after dynamical decoupling. After an initial $\pi/2$ pulse, a variable number of XY16 dynamical decoupling cycles with a fixed time $\tau = 12.5\ \mathrm{ms}$ between $\pi$ pulses are used to offset the  reversible dephasing. The phase of the final $\pi/2$ pulse is chosen to be either 0 or $\pi$, and subtracting the population difference in these two cases provides the coherence contrast. The contrast decay is fitted to obtain $T_2 = 12.6(1)\,\mathrm{s}$. \textbf{d,} Randomized benchmarking of the global single-qubit gate fidelity. For each number of Clifford gates, 60 different random gate strings of this length are applied, after which the overall inverse of the string is applied. For a given gate string length, each translucent marker of a given color represents the return probability for a string of gates while the solid green markers indicate the averaged return probability over the 60 different strings. The inset lists all the colors used to indicate the 60 random gate strings for a given length. The decay of the final population to $1/2$ is fitted to $(1-d)^N$ and $F_c = 1-d/2$ represents the average Clifford gate fidelity.} 
    \vspace{-0.5cm}
    \label{fig:mt-microwave}
\end{figure*}

Moreover, we accurately characterize the imaging survival probability, without assuming any parameters, by performing 80-ms repeated imaging up to 1,000 times, after which ${\sim}90$\% of initially loaded atoms still survive (Fig.~\ref{Fig3}b). This corresponds to a steady-state imaging survival probability of $99.98952(1)\%$, mostly limited by vacuum lifetime. This, to the best of our knowledge, surpasses prior studies reporting record steady-state imaging survival using single alkaline-earth metal~\cite{covey_2000-times_2019} and alkali-metal~\cite{blodgett_imaging_parsed_2023} atoms in optical tweezers.
These results, and the uniformity of imaging survival across the array (Ext. Data Fig.~\ref{fig:survival-ed}a), enable low-loss, high-fidelity detection of single atoms in large-scale arrays, crucial components for the practical use of the system.

\vspace{0.25cm}
\noindent\textbf{Qubit coherence}\\
Key to recent progress in quantum computing and metrology with neutral atoms is the ability to encode a qubit in long-lived states of an atom, such as hyperfine states~\cite{bluvstein_quantum_2022,graham_multi-qubit_2022}, nuclear spin states~\cite{barnes_assembly_2022,huie_repetitive_2023,lis_midcircuit_2023,norcia_midcircuit_2023}, or optical clock states~\cite{norcia_seconds-scale_2019,madjarov_atomic-array_2019}. In cesium atoms, the hyperfine ground states ($\ket{F=3, \ m_F=0}  \equiv \ket{0}$ and $\ket{F=4, \ m_F=0}  \equiv \ket{1}$) provide such a subspace for storing quantum information (see Methods for state preparation and readout procedures). Furthermore, entanglement via Rydberg interactions can be readily transferred to this qubit to realize high-fidelity two-qubit gates~\cite{evered_high-fidelity_2023}. We demonstrate the storage and manipulation of quantum information in a large-scale atom array by measuring the coherence time and global single-qubit gate fidelity using a microwave horn to drive the hyperfine transition (Fig.~\ref{fig:mt-microwave}a). For microwave operation we adiabatically ramp down tweezers to a depth of $k_B \times 55 \ \mathrm{\upmu K}$.

Preserving the coherence of a quantum system as it is scaled up is a known challenge across platforms for quantum computing and simulation~\cite{kjaergaard_superconducting_2020}. This difficulty persists even for neutral atoms, albeit at a lower level, due to residual interactions with a noisy and inhomogeneous electromagnetic environment, particularly with the tweezers themselves. Thus, we choose to trap in far-off-resonant optical tweezers to preserve coherence, since at constant trap depth the differential light shift of the hyperfine qubit decreases as $1/\Delta_\mathrm{tweezer}$ and the scattering rate as $1/\Delta_\mathrm{tweezer}^3$, where $\Delta_\mathrm{tweezer}$ is the tweezer laser detuning relative to the dominant electronic transition~\cite{kuhr_analysis_2005,ozeri_hyperfine_2005,ozeri_errors_2007}. We indeed observe long coherence times, measuring a depolarization time of $T_1 = 119(1)$~s (Ext. Data Fig.~\ref{fig:uw-ed}d), and an array-averaged ensemble dephasing time of $T_2^* = 14.0(1)$~ms (Fig.~\ref{fig:mt-microwave}b), limited by trap depth inhomogeneity. Measured site-by-site, the dephasing time is $T_2^{*(\mathrm{site})} = 25.5$~ms, consistent with being limited by an atomic temperature of ${\sim}4.3\ \mathrm{\upmu K}$ during microwave operation~\cite{kuhr_analysis_2005}. In Ext. Data Fig.~\ref{fig:site-resolved-coherence-ed} and in the Methods we present and discuss site-resolved qubit coherence data.

The dephasing can be further mitigated by dynamical decoupling. By applying cycles of XY16 sequences~\cite{gullion_new_1990} with a period of 12.5 ms between $\pi$ pulses, the measured dephasing time is $T_2 = 12.6(1)$~s, a new benchmark for the coherence time of an array of hyperfine qubits in optical tweezers~\cite{bluvstein_quantum_2022,graham_multi-qubit_2022} (Fig.~\ref{fig:mt-microwave}c). In addition, we investigate in Ext. Data Fig.~\ref{fig:uw-ed}g the coherence time at different trap depths, yielding notably $T_2 = 3.19(5)$ s at the full trap depth of $k_B \times 0.18$ mK. Although lower, this result also surpasses previous known results with hyperfine qubits in a tweezer array.

Finally, we determine single-qubit gate fidelities through global randomized benchmarking~\cite{knill_randomized_2008,xia_randomized_2015,nikolov_randomized_2023}. To compensate for the inhomogeneous Rabi frequency across the array, we use the SCROFULOUS composite pulse~\cite{cummins_tackling_2003}.  We apply gates sampled from the Clifford group $C_{1}$, followed by an inverse operation, and measure the final population in $\ket{1}$ (Fig.~\ref{fig:mt-microwave}d). Fitting the decay as the number of gates increases yields an average Clifford gate fidelity $F_c = 0.999834(2)$, limited by phase noise in our system likely due to magnetic field noise (Methods). This could be readily addressed by upgrading the current sources driving the magnetic field coils or by operating at MHz-scale Rabi rates with optical Raman transitions~\cite{levine_dispersive_2022}.

\vspace{0.25cm}
\noindent\textbf{Towards zone-based quantum computing --- coherent long-distance transport and atom transfer}\\
We now focus more specifically on the practical implementation of a quantum computer, as it is a flagship application of our work and because it demands the most sophisticated toolbox of aforementioned use cases.
Universal quantum computing requires single-qubit and two-qubit gates which have been implemented either through single-site addressing~\cite{graham_multi-qubit_2022} or a zone-based architecture~\cite{bluvstein_quantum_2022,bluvstein_logical_2024}.
Zone-based architectures leverage the ability to dynamically move atoms in a coherence-preserving manner~\cite{beugnon_two-dimensional_2007,bluvstein_quantum_2022}, enabling long-range, non-local connectivity, which allows for less stringent quantum error correction bounds~\cite{bravyi_tradeoffs_2010}. This architecture also provides a path for mid-circuit readout~\cite{bluvstein_logical_2024}. We depict a possible zone layout in Ext. Data Fig.~\ref{fig:transfer-transport-ed}a and the Supplementary Information (SI), which includes a storage zone large enough for more than 6,100 atoms. We do not foresee challenges in creating the zones themselves, e.g., Rydberg-based two-qubit gates should be feasible in a large interaction zone for more than 500 gates in parallel with state-of-the-art fidelities (SI §IV).

However, coherence-preserving transport between storage and adjacent interaction or readout zones might require covering large distances of $\sim$500 $\upmu$m. While moving atoms using acousto-optic deflectors (AODs) is now a well-established practice to resort them into a deterministic configuration~\cite{barredo_atom-by-atom_2016,endres_atom-by-atom_2016} or to transport them coherently~\cite{bluvstein_quantum_2022,shaw_multi-ensemble_2024,bluvstein_logical_2024,finkelstein_universal_2024}, this distance is significantly further than previously demonstrated distances for single-atom transport with tweezers~\cite{bluvstein_quantum_2022, bluvstein_logical_2024}. Furthermore, transferring atoms between dynamic (AOD-generated) and static (SLM-generated) tweezers requires precise relative alignment, conceivably challenging in our system due to the high laser power or potential for worsening aberrations over the large FOV.

\begin{figure}[t!]
     \centering
     \includegraphics[width=89mm]{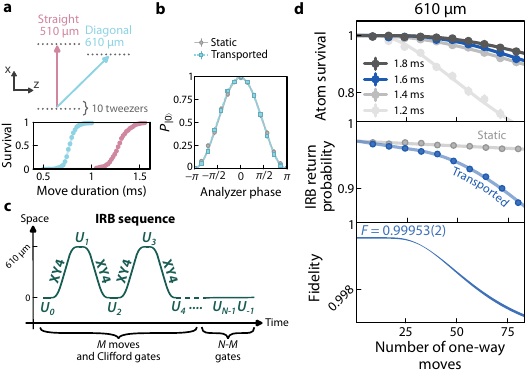}
        \caption{\textbf{Long-distance and high-fidelity coherent transport.}  \textbf{a,} Schematic and atom survival for a diagonal (blue) or straight (pink) move for ten tweezers (with depth $k_B \times 0.28$ mK) spaced by ${\sim}10.6 \ \upmu$m. Despite being shorter, a straight move needs to be executed more slowly than a diagonal one due to cylindrical lensing. \textbf{b,} Coherence of an atom after being transported diagonally 610 $\upmu$m (blue) in $1.6$ ms or held stationary (gray). \textbf{c,} IRB sequence used to benchmark the move fidelity. Random Clifford gates are interleaved between each of the $M (< N)$ moves, with the total number of gates $N$ constant. \textbf{d,} Benchmarking results for repeated 610 $\upmu$m diagonal moves. Top: atom survival for varied times, fitted to a clipped Boltzmann distribution (Methods). $1.6$-ms moves are used for the middle and bottom panels. Middle panel: IRB return probability for static and transported atoms. Curves are fits that include coherence and atom losses (Methods). Bottom panel: average instantaneous transport fidelity after a given number of moves, fitted from the IRB return probability (Methods). The curve width represents the 68\% confidence interval. The instantaneous fidelity of $99.953(2)\%$ is constant for the first ${\sim}30$ moves.}
    \vspace{-0.5cm}
    \label{fig:transport}
\end{figure}

\begin{figure*}[t!]
    \centering
    \includegraphics[width=150mm]{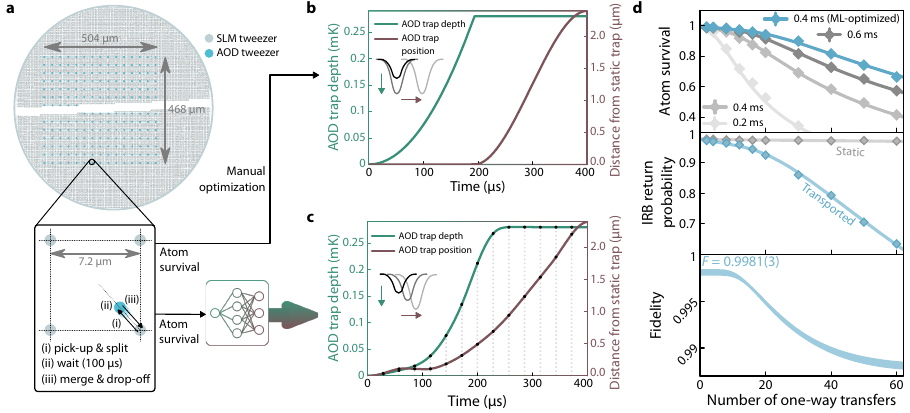}
        \caption{\textbf{Large-scale high-fidelity coherent transfer between static and dynamic traps.} \textbf{a,} Layout of the transfer experiment showing 195 dynamic AOD traps (bright blue) overlapped with 1061-nm static SLM traps (pale blue). Atoms are repeatedly picked up and moved away by 2.4~$\upmu$m, then held for 100~$\upmu$s. During this time, the SLM traps are turned off to ensure that atoms left behind in SLM traps are removed (this way, atom survival correctly corresponds to a successful pick-up and drop-off). Then SLM traps are subsequently turned on, and atoms held in AOD tweezers are moved back and dropped off into the SLM traps. For IRB data shown in (d), gates are interleaved between each round-trip transfer. A pick-up and split-move operation (or equivalently a merge-move and drop-off operation) are considered a `one-way transfer'. \textbf{b,} Best hand-optimized trajectory for trap transfer (Methods), using a quadratic depth profile and a constant jerk movement. Here we implement the pick-up and the tweezer separation move in sequence, without overlap. \textbf{c,} To speed up atom transfer between static and dynamic traps while preserving high survival, we optimize, via machine learning, a trajectory where dynamic AOD traps are simultaneously ramped and moved. The dashed lines and black dots represent the values that are optimized by the algorithm. \textbf{d,} Top panel: atom survival as a function of the number of repeated one-way transfers for various one-way `pick-up \& split' total durations. A $400\mu$s trajectory is optimized through machine learning. Middle panel: return probability after IRB for the machine-learning optimized trajectory. Bottom panel: extracted instantaneous fidelity of a coherent one-way transfer as a function of the number of previous one-way transfers.}
    \vspace{-0.5cm}
    \label{fig:transfer}
\end{figure*}

Thus, we investigate the feasibility of coherence-preserving transport and SLM-AOD trap transfer over larger length scales. First, isolating challenges with the coherent transport operation, we load atoms directly into ten AOD-generated tweezers and characterize coherent moves up to $\sim$610 $\upmu$m (Fig.~\ref{fig:transport}, top section of Ext. Data Fig.~\ref{fig:transfer-transport-ed}). Second, we assess the viability of large-scale parallel AOD-SLM trap transfers with 195 AOD-generated tweezers that span a square of 504 $\upmu$m $\times$ 468 $\upmu$m (Fig.~\ref{fig:transfer}). As an outlook, we demonstrate a proof-of-principle combination of these techniques in a large-scale static array (although in a different trap layout) moving a 2D array of 47 atoms over 375 $\upmu$m,  a distance comparable to predicted zone spacings in our system (Ext. Data Fig.~\ref{fig:transfer-transport-ed}e, f). For all operations, we utilize the most wide-band commercially available AODs at near-infrared wavelengths, which cover up to $500-600 \ \upmu$m along one axis for the optical parameters used here (Methods).  

In probing atom survival as a function of long-distance movement speeds, we find that the speed of transport is strongly constrained by cylindrical lensing---an effect that occurs when the AOD frequency is rapidly swept~\cite{dickson_optical_1972}---which becomes increasingly deleterious as the AOD field of view is increased (SI III.A). Notably, using a pair of crossed AODs for diagonal transport converts cylindrical lensing into spherical lensing, enabling significantly faster movement (Fig.~\ref{fig:transport}a).
With diagonal moves, we first demonstrate in Fig.~\ref{fig:transport}b negligible loss of coherence for atoms transported by 610 $\upmu$m in 1.6 ms. We suppress dephasing with one XY4 dynamical decoupling sequence per move. 

Realistic applications of coherent transport involve multiple consecutive moves. Therefore, we characterize the fidelity of the \textit{quantum channel} defined by coherent transport through interleaved randomized benchmarking~\cite{magesan_efficient_2012} (IRB, Fig.~\ref{fig:transport}c, Methods). To the best of our knowledge, such a quantitative characterization of transport fidelity in neutral atoms has not been previously demonstrated. To maximize the dephasing cancellation, we apply dynamical decoupling in a transformed Clifford frame (Methods).

We perform this benchmarking technique for a distance of $610 \ \upmu$m (Fig.~\ref{fig:transport}d), with diagonal moves. We first measure the survival probability of an atom in a tweezer at the end of the sequence for different move durations (top panel). For a 1.6-ms move using $k_B \times 0.28$ mK-deep traps, we then characterize the return probability to the initial quantum state after IRB as a function of the number of moves (middle panel). Other distances, trap depths and move times are shown in Ext. Data Fig.~\ref{fig:transfer-transport-ed}.

The resulting IRB return probability data is non-exponential in the number of moves, because at large numbers of moves, trap losses become dominant and the fidelity for the transport channel depends on the number of previously executed moves. This motivates defining an \textit{instantaneous} fidelity; i.e., the fidelity of the transport channel after a certain number of previous moves (Methods), shown in the bottom panel of Fig.~\ref{fig:transport}d. The instantaneous fidelity approaches a constant value of $99.953(2)\%$  for small numbers of one-way moves ($\lesssim 30$), where losses are the sub-dominant error. This regime is most relevant for QEC, since data qubits and ancilla qubits can, in principle, be swapped every few layers of gates~\cite{chow_circuit-based_2024_new}.

We then move on to characterizing the atom transfer between static and dynamic tweezers. We demonstrate that these operations proceed without the emergence of unexpected technical challenges by performing high-fidelity parallel AOD-SLM transfer across the full field of view of the AOD (Fig.~\ref{fig:transfer}).

We use 195 AOD tweezers spread across 504 $\upmu$m $\times$ 468 $\upmu$m (Fig.~\ref{fig:transfer}a) to perform and characterize the repeated transfer procedure, post-selected on initially filled SLM sites. As with coherent transport benchmarking, we evaluate the transfer fidelity as a function of the number of one-way transfers through IRB (Fig.~\ref{fig:transfer}d). To execute faster (or higher-fidelity transfers at a given duration), we propose and implement a trajectory where AOD ramp-and-move operations are simultaneously optimized with machine learning techniques to maximize survival (Fig.~\ref{fig:transfer}c, Methods). Compared with our manually optimized trajectory (Fig.~\ref{fig:transfer}b), this technique yields significantly higher atomic survival, and enables a one-way transfer fidelity of 99.81(3)\% for $\lesssim 12$ transfers (compare blue and gray data in Fig.~\ref{fig:transfer}d, top panel). 

In the future, such machine learning techniques could also be used to optimize combined pick-up and transport, where we find a fidelity of $99.87(1)\%$ for the first ${\sim}12$  operations at the chosen timescales with manually optimized methods (Methods, Ext. Data Fig.~\ref{fig:transfer-transport-ed}f).

Finally, to cover the full extent of the array, we envision utilizing multiple pairs of crossed AODs, with the demonstrated long-distance transport allowing overlap between adjacent AOD-pair controlled regions (SI Fig.~2). With the layout presented in Ext. Data Fig. \ref{fig:transfer-transport-ed}a and the SI, four such regions would be necessary. Alternatively, additional scanning techniques (e.g., fast-scanning mirrors) can be used to position the field of view of a single pair of crossed AODs across the full array iteratively. 

Such techniques are also applicable to initial rearrangement of atoms in the storage zone. For example, by implementing a parallel assembly algorithm\cite{tian_parallel_2023,wang_accelerating_2023} in four quadrants (SI §II), with estimates for relevant timings based on simulation, data, and previous experiments (SI Table I), we expect that we can sort the array in parallel in $\sim$137 ms or sequentially quadrant-by-quadrant in $\sim$522 ms.

\vspace{0.25cm}
\noindent\textbf{Conclusion and outlook}\\
We have shown scaling of neutral atom qubit numbers in optical tweezers to more than 6,100. We simultaneously achieve high imaging survival and fidelity as well as a long room-temperature vacuum-limited lifetime. We find record coherence times in alkali-metal atom tweezer arrays and a high global single-qubit gate fidelity, limited by technical noise. Further, we additionally characterize the fidelity of quantum transport channels for moves and trap transfer at relevant length scales, utilizing randomized benchmarking.

Our results usher in a new generation of neutral atom quantum processors based on several thousand qubits, particularly relevant for QEC~\cite{bravyi_high-threshold_2024,xu_constant-overhead_2024}. Additionally, large-scale programmable devices enabling advances in quantum metrology~\cite{rosenband_exponential_2013,ludlow_optical_2015,norcia_seconds-scale_2019,madjarov_atomic-array_2019,young_half-minute-scale_2020,finkelstein_universal_2024} and simulation~\cite{haferkamp_closing_2020, orourke_entanglement_2023, julia-farre_amorphous_2024} are made accessible through this work. For example, our platform---with the demonstrated qubit numbers---could be used for verifiable quantum advantage with low-depth evolution~\cite{haferkamp_closing_2020,ringbauer_verifiable_2025}. Tweezer clocks could be scaled using near-infrared, high-power tweezers for loading and imaging~\cite{tao_high-fidelity_2024} before transferring atoms to magic-wavelength traps for clock operation~\cite{norcia_seconds-scale_2019,madjarov_atomic-array_2019,young_half-minute-scale_2020,finkelstein_universal_2024}.  We also foresee applications in quantum simulation for problems where boundary effects play an important role~\cite{bernien_probing_2017,browaeys_many-body_2020,ebadi_quantum_2021,scholl_quantum_2021, orourke_entanglement_2023}, which can be minimized with the large system sizes demonstrated here.

Finally, our work indicates that further scaling of the optical tweezer array platform to tens of thousands of trapped atoms should be achievable with current technology, while essentially preserving high-fidelity control. In our present apparatus, several factors limit the number of sites. One limitation is the finite number of pixels of each SLM (reducing the diffraction efficiency as the array size is increased), along with reduced SLM diffraction efficiency at higher incident laser powers. By using available higher resolution SLMs, and by exploring techniques with higher pixel modulation depth~\cite{moreno_diffraction_2020}, we hope to utilize both power and field of view more efficiently. 

Furthermore, we observe worsening optical aberrations at tweezer powers greater than that in the present study due to thermal heating of the objective. This is the main limitation on atom number for the results in this work, even after aberrations were mitigated using the SLM (Methods). This constraint could be circumvented by utilizing an objective with a housing material that retains less heat or with integrated cooling strategies. Such upgrades should allow us to almost double the number of tweezers that we create using two fiber amplifiers. We further anticipate the potential to switch from polarization combination to wavelength-based array combination, opening up further avenues for increasing tweezer number with similar techniques to those utilized in this work. Atom numbers may further be increased in our array with the same number of tweezers by utilizing enhanced loading~\cite{brown_gray-molasses_2019} or re-loading techniques~\cite{shaw_dark-state_2023,norcia_iterative_2024_mod}. Already in the near term, we expect to increase the number of atomic qubits to over ten thousand with the current system using a subset of these techniques.

\begin{acknowledgements}
We acknowledge insightful discussions with, and feedback from, Adam Shaw, Harry Levine, Zunqi Li, Ran Finkelstein, Pascal Scholl, Joonhee Choi, Dolev Bluvstein, and Soonwon Choi. We acknowledge support from the Gordon and Betty Moore Foundation (Grant GBMF11562), the Weston Havens Foundation, the Institute for Quantum Information and Matter, an NSF Physics Frontiers Center (NSF Grant PHY-2317110), the NSF QLCI program (2016245), the NSF CAREER award (1753386),  the Army Research Office MURI program (W911NF2010136), the U.S. Department of Energy (DE-SC0021951), the DARPA ONISQ program (W911NF2010021), the Air Force Office for Scientific Research Young Investigator Program (FA9550-19-1-0044), and the Heising-Simons Foundation (2024-4852). Support is also acknowledged from the U.S. Department of Energy, Office of Science, National Quantum Information Science Research Centers, Quantum Systems Accelerator.
H.J.M. acknowledges support from the NSF Graduate Research Fellowship Program under Grant No. 2139433. K.H.L. acknowledges support from the AWS-Quantum postdoctoral fellowship and the NUS Development Grant AY2023/2024.

\end{acknowledgements}

\section*{Data availability}
The data and codes that support the findings of this study are available from the corresponding author upon request.

\section*{Author contributions}
H.J.M. and M.E. conceived the idea and experiment. H.J.M., G.N., E.B., K.H.L., and X.L. performed numerical simulations and contributed to the experimental setup. H.J.M., G.N., E.B., and K.H.L. performed the experiments. H.J.M., G.N., and E.B. analyzed the data. All authors contributed to the manuscript. M.E. supervised this project.

\section*{Competing interests}
The authors declare no competing interests.

\putbib
\end{bibunit}

\begin{bibunit}

\setcounter{figure}{0}
\captionsetup[figure]{labelfont={bf},name={Extended Data Fig.},labelsep=bar,justification=raggedright,font=small}

\begin{figure*}[hbt!]
	\centering
	\includegraphics[width=180mm]{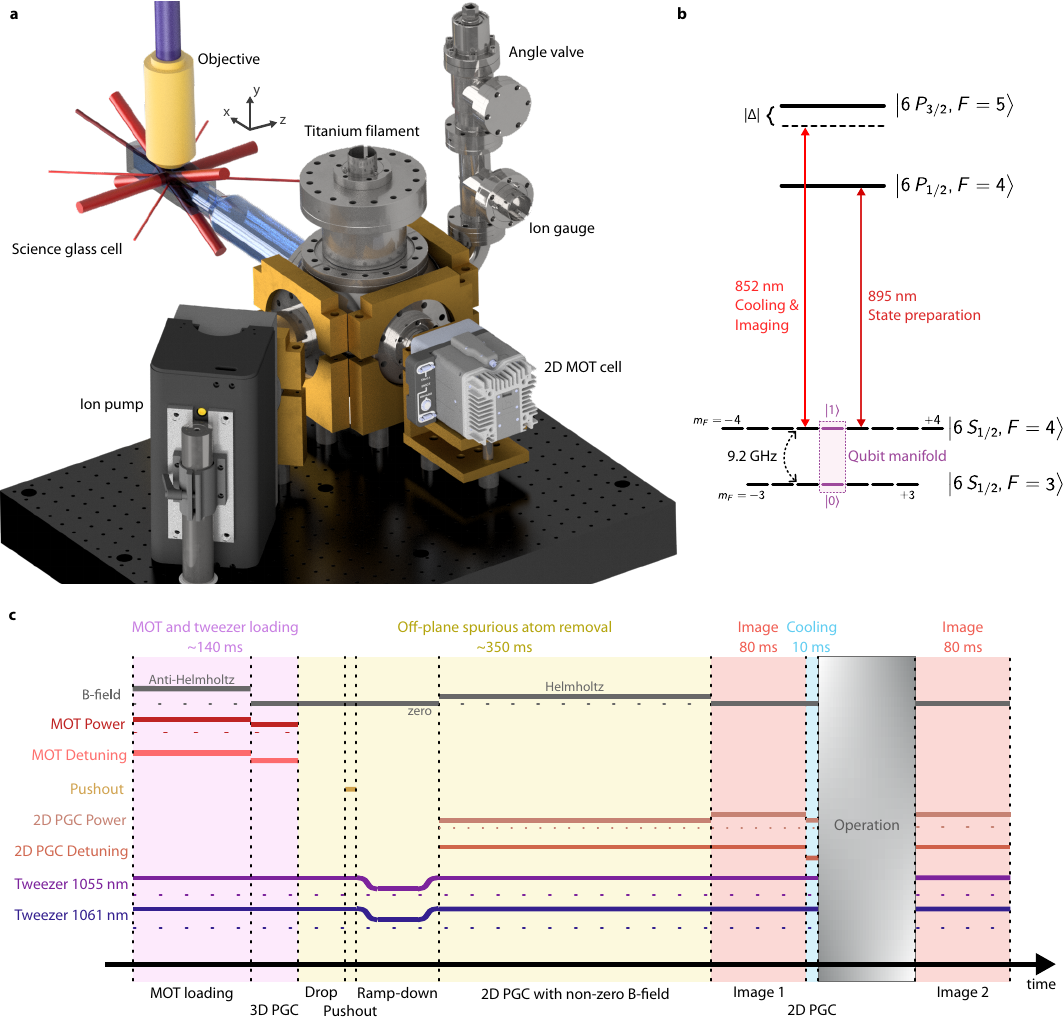}
	\caption{\textbf{Experiment apparatus and sequence.} \textbf{a,} Simplified view of the vacuum chamber. The 2D MOT cell (Infleqtion CASC) containing an electrically heated cesium dispenser, shown inside its integrated photonics assembly. It is attached to a stainless steel vacuum chamber on which an ion pump is mounted. We further use two titanium sublimation pumps (one mounted from the top, as shown, and one mounted from the bottom, not visible), sputtering almost the entire surface area of the chamber, except the rectangular part of the science glass cell and the ion pump. We use the following conventions for the laser beams: thick red for MOT beams, thin red for PGC beams, dark red (along $\hat{x}$) for state preparation beam, and purple for tweezer beam. \textbf{b,} Summary of the relevant states and transitions used in this work. \textbf{c,} Summary of a typical experimental sequence, as described in the Methods.}
	\label{ext_fig1}
\end{figure*}

\begin{figure*}[hbt!]
	\centering
	\includegraphics[width=91mm]{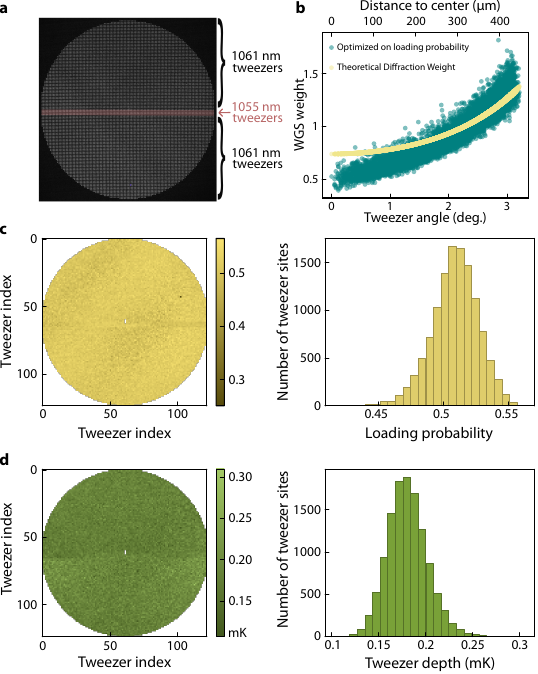}
	\caption{\textbf{Tweezer uniformity details.} \textbf{a,} The tweezers created by two fiber amplifiers are labeled on the averaged atomic image shown in Fig.~\ref{Fig1}b. We create 11,513 (488) tweezers with laser light at 1061 nm (1055 nm), for a total of 12,001 tweezers. The 1055-nm tweezers fill the gap created by the spatial filtering of the 0\textsuperscript{th} order in the 1061-nm tweezer pathway, as described further in the tweezer generation section. \textbf{b}, The WGS weights given to tweezers during the tweezer homogenization procedure, as a function of angular distance from the 0\textsuperscript{th}-order reflection off the SLM, with the physical distance this corresponds to given our optical setup shown on the upper axis. In teal is plotted the weights obtained after the tweezer depths are uniformized based on loading probability. In yellow is shown the weight compensation that would be expected based on diffraction efficiency calculations assuming blazed gratings are utilized for displacement. The weight increases with larger angle in order to compensate for the diminishing diffraction efficiency as a function of tweezer distance to the center. This additionally informs our decision to create a circularly shaped array. \textbf{c,} The per-site loading probability array map and its histogram. We feedback on the WGS trap depths based on the loading rate per site to uniformize the trap depth. We see an average loading probability per site of 51.2\% with a relative standard deviation of 3.4\%. The lowest loading probability is 25.1\% for one tweezer, which is the only tweezer not shown in the histogram, but included in the quoted average. This tweezer, however, does not exhibit a significant difference in imaging survival probability, coherence time, or single-qubit gate fidelity (see Ext. Data Fig.~5a and Ext. Data Fig.~8). Therefore, this tweezer site could also be utilized in the experiment. Note that three tweezers in the array are excluded for the data shown in this work, since they are affected by leakage from the 0\textsuperscript{th} order of the SLM on the 1061-nm tweezer pathway that is not completely extinguished via the spatial filtering, resulting in 11,998 usable sites out of 12,001 generated sites. We note that quantum computation can be done despite defects in an array, and that thus one can choose to avoid a small number of traps~\cite{strikis_quantum_2023}. \textbf{d,} The per-site tweezer depth map and its histogram. This is obtained by measuring the differential light shift on $F=4 \leftrightarrow F'= 4$ $D2$ transition~\cite{levine_quantum_2021}. We see an average trap depth of $k_B \times0.18(2)$ mK with a standard deviation of 11.4\% across the sites.}
	\label{fig:loading-ed}
\end{figure*}

\begin{figure*}[hbt!]

    \includegraphics[width=152mm]{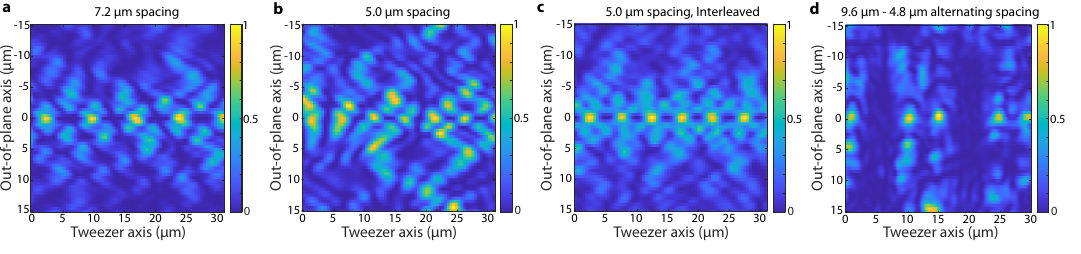}
    \caption{\textbf{Tweezer spacing details.} Calculation of an out-of-plane intensity profile from a section of a 122 by 122 site tweezer array at 1061 nm, whose phase pattern is generated by a WGS algorithm. The tweezer axis (x-axis in plots) is centered along a selected row in the array and the out-of-plane axis (y-axis in plots) is perpendicular to the focal plane of the tweezers, along the direction of light propagation. The focal plane for the tweezers is at 0 $\upmu$m, and we simulate for different spacing between the tweezers: \textbf{a,} 7.2 $\upmu$m and \textbf{b,} 5.0 $\upmu$m. We also show in \textbf{c,} the case for which 5.0 $\upmu$m spacing is achieved by alternating traps generated with two different lasers such that they do not interfere by using for example orthogonal polarization or sufficiently different wavelengths. One could imagine using such an interleaved configuration to achieve tighter tweezer spacing without being limited by out-of-plane interference, in order to increase the number of atoms within the field of view. In \textbf{d,} we show the out-of-plane interference for the case of alternating spacing between tweezers of 9.6~$\upmu$m and 4.8 $\upmu$m as is used in Ext. Data Fig.~\ref{fig:transfer-transport-ed}e,f. The color scale for each case is normalized by the highest intensity in the simulated slice.}
     \label{fig:out-of-plane-ed}
\end{figure*}

\begin{figure*}[hbt!]
	\centering
	\includegraphics[width=152mm]{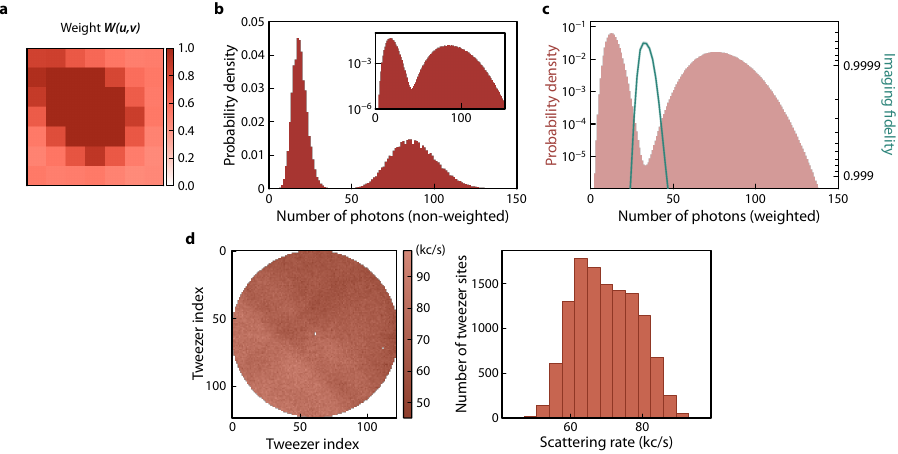}
	\caption{\textbf{Imaging characterization.} \textbf{a,} Weight function $W(u,v)$ applied to each pixel of the $7\times 7$ square-pixel box around each site. Here, $u$ and $v$ refer to the camera pixel coordinates centered on a given site. \textbf{b,} Imaging histogram obtained by summing the number of photons in the $7\times 7$ square-pixel box around each site, without any weights. \textbf{c,} Imaging fidelity as the binarization threshold is displaced from its optimal position. \textbf{d,} Map and histogram of the scattering rate per site across the tweezer array.}
	\label{fig:imaging-ed}
\end{figure*}

\begin{figure*}[hbt!]
	\centering
	\includegraphics[width=90mm]{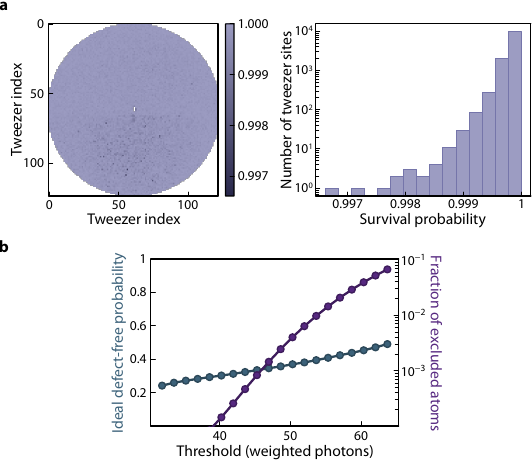}
	\caption{\textbf{Imaging survival details.} \textbf{a,} Map and histogram of the imaging survival probability per site across the tweezer array, as characterized using the three image analysis of data from 16,000 iterations. Note that the vertical axis of the histogram figure is plotted on a log scale. The mean of site-resolved imaging survival probability is 99.985\%, and the minimum value 99.66\%. \textbf{b,} Predicted upper bound on the probability of detecting a defect-free array after an ideal rearrangement sequence (estimated as $p(1|1)^n$ where $n$ is the number of atoms in the first image), limited by imaging survival and false positives. The threshold in the first image can be displaced to reduce false positives, at the cost of excluding some atoms. Note that we may ignore the issue of false negatives in the first image, since we can always physically eject residual atoms in sites that are determined to be negative.}
	\label{fig:survival-ed}
\end{figure*}

\begin{figure*}[hbt!]
    \centering
    \includegraphics[width=72mm]{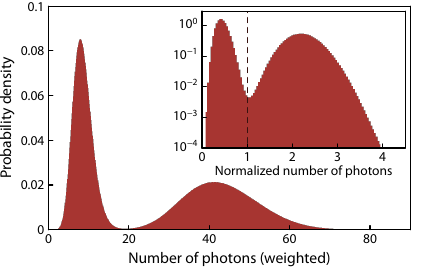}
    \caption{\textbf{Imaging in 20 ms.} Imaging histogram obtained with an imaging time of 20 ms. The weight function is the same as the one shown in Ext. Fig.~3a. Using the model-free imaging characterization, we find an imaging fidelity of 99.9571(4)\% and a survival probability of 99.176(1)\%. Inset: log-scale histogram where the number of photons for each site is rescaled by the threshold for this site.}
    \label{fig:fidelity-fastimg-ed}
\end{figure*}

\begin{figure*}[hbt!]
	\centering
	\includegraphics[width=138.5mm]{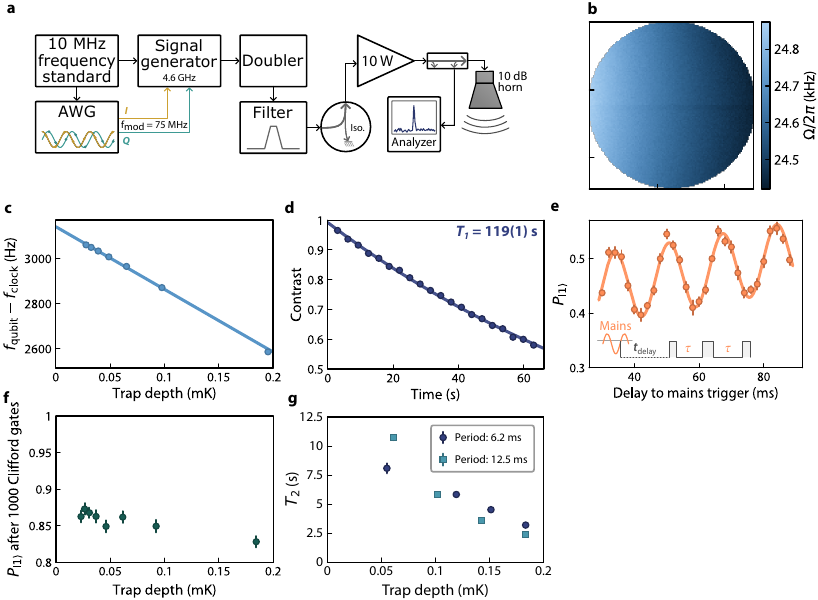}
	\caption{\textbf{Characteristics of microwave-driven qubits.} \textbf{a,} Schematic of the setup used to drive the hyperfine qubit. \textbf{b,} Inhomogeneity of the Rabi frequency across the atom array. The reflection off a vertical metallic breadboard near the vacuum cell creates this spatial gradient orthogonal to the propagation axis of the microwaves. The Rabi frequency standard deviation is 0.5\%. \textbf{c,} Estimation of $\eta$, the ratio of the differential polarizability of the hyperfine qubit to the electronic ground state scalar polarizability. The average qubit frequency is measured by Ramsey interferometry for different trap depths, and the slope is compared with the trap depth inferred from the light shift of the $F = 4 \leftrightarrow F' = 4$ $D2$ transition from its free-space resonance. \textbf{d,} Measurement of the depolarization time $T_1$. Atoms are initially prepared in $\left| 1 \right\rangle$. After a given time, the remaining population in $\left| 1 \right\rangle$ is measured, with or without a $\pi$ pulse before the measurement. The population difference, conditioned by the application of the pulse, constitutes the $T_1$ contrast. This extended $T_1$ time is largely enabled due to the choice of far-detuned traps. \textbf{e,} A spin-echo sequence is employed to probe the 60 Hz phase noise in our system. The free-evolution time of each arm, $\tau$, is set to a half-period of 60 Hz, which enhances the noise. By varying the time $t_{\mathrm{delay}}$ between the line trigger and the spin-echo sequence, we map the periodic noise at 60 Hz to the population in $\left| 1 \right\rangle$. \textbf{f,} The population in $\left| 1 \right\rangle$ after 1,000 random Clifford gates is measured for different trap depths, exhibiting only limited improvement when the trap depth is reduced. In addition to other elements presented in the Methods, this indicates that the single-qubit gate fidelity is likely limited by residual magnetic field noise, which could be readily mitigated by technical improvements (Methods). Error bars indicate estimated 68\% confidence intervals. \textbf{g,} Measurement of the coherence time $T_2$ at different trap depths, for two different periods between $\pi$ pulses. Error bars indicating 68\% confidence intervals are shown when larger than the dot itself. Note that this experiment was performed with the 11,513 tweezers generated by the fiber amplifier at 1061 nm only (Ext. Data Fig.~\ref{fig:loading-ed}a).}
	\label{fig:uw-ed}
\end{figure*}

\begin{figure*}[hbt!]
	\centering
	\includegraphics[width=91mm]{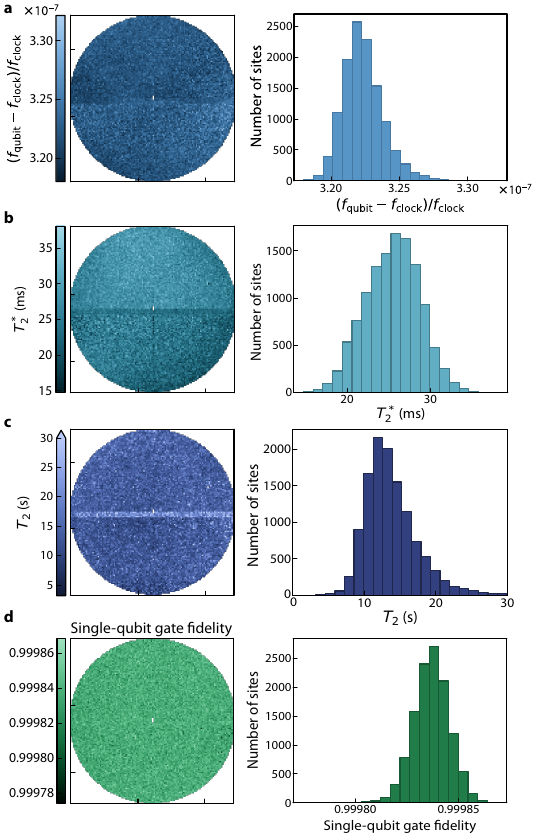}
	\caption{\textbf{Site-resolved coherence metrics.} \textbf{a,} Relative difference of the qubit frequency with the cesium clock frequency $f_{\mathrm{clock}} \equiv 9,192,631,770$ Hz, measured by Ramsey interferometry. The standard deviation is $1.5\times 10^{-9}$, or 14 Hz in absolute value. \textbf{b,} Map and histogram of $T_2^*$ across the atom array. The average uncertainty per site is 1.5 ms. The average $T_2^*$ for sites generated by the fiber amplifier at 1055 nm is 23.2(1) ms, while it is 25.58(3) ms for sites generated by the fiber amplifier at 1061 nm. \textbf{c,} Map and histogram of $T_2$ across the atom array. The average uncertainty per site is 2.8 s. The average $T_2$ for sites generated by the fiber amplifier at 1055 nm is 19.2(4) s, while it is 12.32(6) s for sites generated by the fiber amplifier at 1061 nm. We use averages weighted by the uncertainty on each site, since we observe that the unweighted average results in a bias from the value obtained by global fitting.  \textbf{d,} Map and histogram of single-qubit gate fidelity obtained by global randomized benchmarking. The average gate fidelity is 99.9834(2)\%.}
	\label{fig:site-resolved-coherence-ed}
\end{figure*}

\begin{figure*}[hbt!]
	\centering
	\includegraphics[width=89mm]{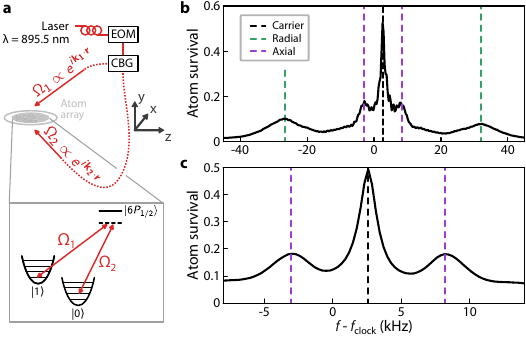}
	\caption{\textbf{Raman sideband spectroscopy.} \textbf{a,} Schematic of the Raman configuration used to address the atomic motion. The amplitude modulation setup and Raman configuration are detailed in the Methods. \textbf{b,} Raman spectroscopy results exhibiting sidebands corresponding to the radial motion (in green) and the axial motion (in purple). We measure an radial trapping frequency of 29.30(4) kHz and an axial trapping frequency of 5.64(3) kHz. The sideband signal is broadened due to inhomogeneities in the array. The measurement is averaged over the 11,513 tweezers created with the 1061 nm light as shown in Ext. Data Fig. \ref{fig:loading-ed}a \textbf{c,} Fine-grained spectroscopy data acquired with a lower Rabi frequency to resolve the axial sideband.}
	\label{fig:raman-sideband-ed}
\end{figure*}

\begin{figure*}[hbt!]
	\centering
	\includegraphics[width=160mm]{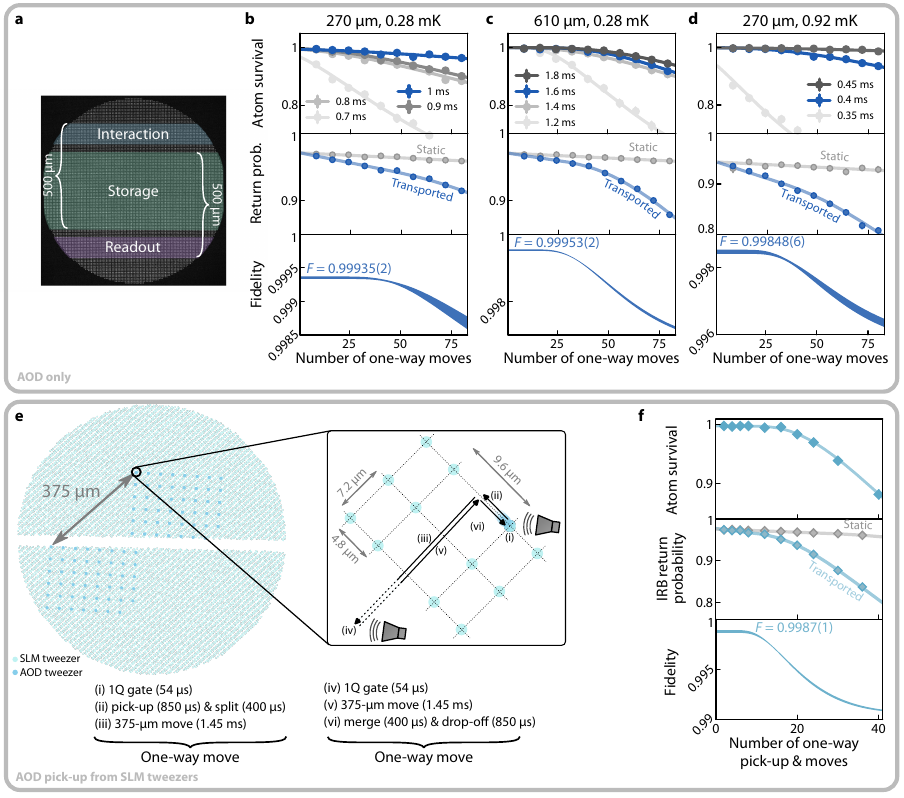}
	\caption{\textbf{Long distance AOD movement and large-scale AOD-SLM trap transfer.} \textbf{[AOD only] a,} Proposed layout of a zone-based universal quantum processor with 6,100 atoms. Atoms anywhere in the storage zone can be transported with AODs to the interaction or readout zones in under 500 $\upmu$m. \textbf{b, c, d,} Results of the randomized benchmarking of transport, for different distances and trap depths, as specified above each subfigure. Similarly to Fig.~\ref{fig:transport}d we present the atomic survival for various move durations (top panel), the IRB return probability for the specific duration highlighted in blue (middle panel), and the extracted instantaneous transport fidelity. The curve width in the bottom panel represents the 68\% confidence interval. \textbf{[AOD and SLM]} \textbf{e,} Schematic representing the configuration and operations used for coherent transfer and transport of atoms using 47 AOD tweezers. The SLM layout (totaling 11,416 sites) alternates narrow column spacings of 4.8 $\upmu$m and wide column spacings of 9.6 $\upmu$m, between which AOD tweezers are moved diagonally (see Methods). Timescales for each operation composing one-way moves are detailed in the figure. \textbf{f,} Results of the randomized benchmarking of coherent transfer and transport, similarly to previous figures.}
	\label{fig:transfer-transport-ed}
\end{figure*}

\FloatBarrier

\section*{Methods}


\subsection*{Vacuum apparatus}
A schematic of our vacuum system is shown in Ext. Data Fig.~\ref{ext_fig1}. After the initial chamber assembly and multi-round baking process, we fire two titanium sublimation pumps (TSPs), mounted such that every surface except the rectangular portion of the glass cell and the interior of the ion pump are covered by line-of-sight sputtering. This creates a vacuum chamber in which essentially every surface is pumping. We do not find it necessary to re-fire the TSPs in order to maintain the vacuum level that we measure. We additionally maintain ultra-high vacuum conditions with an ion pump, connected to the primary chamber via a $45^{\circ}$ elbow joint. The secondary, science chamber consists of a rectangular glass cell (JapanCell) optically bonded to a 24-cm long glass flange (also sputtered by the TSP) that connects to the primary chamber. From lifetime measurements of tweezer trapped atoms (see main text) and collisional cross-sections available in literature~\cite{monroe_very_1990}, we estimate the pressure in the glass cell to be ${\sim}7\times 10^{-12}$ mbar, consistent with vacuum simulations using the MolFlow program~\cite{kersevan_recent_2019}. 

\subsection*{Tweezer generation}
We utilize light from two fiber amplifiers, at 1061 nm (Azurlight Systems) and 1055 nm (Precilasers) to create the optical tweezers through an objective (Special Optics) with $\mathrm{NA}=0.65$ at the trapping wavelengths ($\mathrm{NA}=0.55$ at the imaging wavelength of 852 nm) and a field of view of 1.5 mm. The tweezers are imprinted onto the light in each pathway by a Meadowlark phase-only Liquid Crystal on Silicon Spatial Light Modulator (SLM) that is water cooled to maintain a temperature of 22 $^{\circ}$C. On each path, there are two $4f$ telescopes utilized to map the SLM phase pattern onto the back focal plane of the objective, which subsequently focuses the tweezers into the vacuum cell as shown in Fig.~\ref{Fig1}c. In the first focal plane after the SLM, we perform spatial filtering on the two paths in order to remove the 0\textsuperscript{th} order and reflect the 1\textsuperscript{st}-order diffracted light from the SLM. On the 1061-nm path we use two D-mirrors spaced by a few hundred microns, and on the 1055-nm path we use a mirror with a manufactured 300-$\upmu$m hole as spatial filters to separate 0\textsuperscript{th}-order light from the tweezer light. The 1055-nm tweezers are essentially used to fill the gap between two halves of the array created by the 1061-nm tweezers (Ext. Data Fig.~\ref{fig:loading-ed}a), although we anticipate increasing the number of tweezers created with this path after implementing the objective heat-dissipation strategies as described in the discussion and outlook section. We currently use 120 W of power from the 1061-nm fiber amplifier, and around 10 W of power from the 1055-nm fiber amplifier to create the tweezers. On the 1061-nm path after all the optical elements, we estimate that only around 35-40 W of the total power reaches the objective and, given measurements of trap parameters, that we have ${\sim}1.4$ mW per tweezer. At low optical power, we estimate a ratio between the incoming power and the light diffracted into the 1\textsuperscript{st} order of the SLM of around 65\% into the full array, and at full optical power, we estimate a diffraction efficiency of around 45\%, even after optimizing the SLM global calibration at high power. We leave further improvement to future work.

While one would like to separate the 1\textsuperscript{st}-order hologram phase pattern and 0\textsuperscript{th}-order reflection in a more convenient manner, the largest angular separation that is possible between the 0\textsuperscript{th} and 1\textsuperscript{st} orders of the SLM, as determined by the SLM pixel size, would not separate the large tweezer array from the 0\textsuperscript{th} order, due to the large angular distribution of the tweezers. Furthermore, the diffraction efficiency of the SLM into the 1\textsuperscript{st} order decreases with increasing separation from the 0\textsuperscript{th} order. Therefore, it is the most power-efficient choice to center the tweezers around the 0\textsuperscript{th} order, and to filter it at the first focal plane after the SLM. This decreasing diffraction efficiency with increasing distance from the 0\textsuperscript{th} order, at the center of the array, informs our choice of a circular tweezer array. We highlight the development of these techniques of 0\textsuperscript{th} order filtering as uniquely necessary for a large-scale array.

The SLM phase patterns are optimized with a weighted Gerchberg-Saxton (WGS) algorithm~\cite{kim_gerchberg-saxton_2019, kim_large-scale_2019} to create a tweezer array that we uniformize through a multi-step process, first adjusting weights in the algorithm based on photon count on a CCD camera that images the tweezers~\cite{nogrette_single-atom_2014}, and secondly adjusting weights based on the loading probability of each site in the atomic array with a variable gain feedback, as demonstrated on smaller arrays in previously developed schemes~\cite{schymik_situ_2022}. We implement around 5 iterations of each step in order to achieve the loading and survival probabilities that are shown in Ext. Data Figs.~\ref{fig:loading-ed}c,~\ref{fig:survival-ed}a. The WGS goal weight $W_i$ on each tweezer for the $i^{\textrm{th}}$ iteration is given by 
\begin{equation*}
W_i = \frac{1-G(1-\sqrt{H_i})}{\langle W_i \rangle},
\end{equation*}
normalized by the mean weight $\langle W_i \rangle$,
where the height $H_i$ is determined by adjusting the value from the previous iteration using the loading probability per tweezer $P_{\textrm{load}}$, normalized by the average loading probability,
\begin{equation*}
H_i = H_{i-1} \left[1-g \left(1-\frac{P_{\textrm{load}}}{\langle P_{\textrm{load}} \rangle }\right)\right].
\end{equation*}
We choose the weight of the gains $G$ and $g$ in order to reach convergence for the given configuration of tweezers (here we use a value of $0.6$ for each), and additionally add a cap to the allowable values of $H_i$ in order to avoid oscillatory behavior. We show in Ext. Data Fig.~\ref{fig:loading-ed}b, the weights for tweezers for different angular diffraction off of the SLM, obtained after utilizing the loading-based uniformization. We also show the theoretical weights that would be expected based on the inverse of the naive diffraction efficiency calculations for blazed gratings. The diffraction efficiency is given by $ DE = \sinc^2\left(\frac{\pi a x }{ \lambda f}\right) \sinc^2\left(\frac{\pi a y }{ \lambda f}\right) $, where $a$ is the SLM pixel size, $x$ and $y$ are the horizontal and vertical displacements from the 0\textsuperscript{th} order at the tweezer plane, $f$ is the effective focal length of the objective, and $\lambda$ is the trapping wavelength. We expect that some divergence in behavior could be due to angular-dependent transmission in optics in the imaging path.

We furthermore add aberration correction to the SLM phase hologram based on Zernike polynomials~\cite{levine_quantum_2021}. We perform a gradient-descent-type optimization to determine the amplitude of the Zernike polynomial coefficients that maximizes the filling fraction in the array. We iterate between this optimization and 2-3 rounds of loading-based uniformization. 

To align the tweezers created by the two fiber amplifiers in angle, we change the goal configuration for the WGS algorithm. The CCD camera on which we image the tweezers after the vacuum cell provides a helpful reference for this alignment.

\subsection*{Loading single atoms in tweezers}
The typical experimental sequence can be seen in Ext. Data Fig.~\ref{ext_fig1}c. From an atomic beam generated with a two-dimensional magneto-optical trap (2D MOT) of cesium-133 atoms (Infleqtion CASC), we load ${\sim}10^{7}$ atoms in the three-dimensional (3D) MOT in 100 ms using three pairs of counter-propagating beams in each axis and create a ${\sim}1.6$-mm $1/e^2$ diameter MOT cloud. The magnetic field gradient is set to 20 $\mathrm{G\, cm^{-1}}$ with a quadrupole configuration using a pair of coils that is perpendicular to the objective axis. Each beam has a size of 2.5 cm in diameter, detuning of $\Delta=-3.17 \Gamma$ from the bare atom $\ket{6 S_{1/2}, F=4}\leftrightarrow\ket{6 P_{3/2}, F'=5}$ resonant transition (Ext. Data Fig.~\ref{ext_fig1}b), and a total intensity of 10$I_0$ (1.6$I_0$ for repumping beams), where $I_0\approx1.1 \, \mathrm{mW \, cm^{-2}}$ is the saturation intensity of the transition between the stretched states, and $\Gamma \approx 2\pi \times 5.2 \ \mathrm{MHz}$ is the natural linewidth of the $6 P_{3/2}$ electronically excited state~\cite{steck_cesium_2023}. After loading atoms into the 3D MOT, we switch off the quadrupole magnetic field and, at the same time, lower the intensity to $7 I_0$ and detune the laser further to $\Delta=-19.5\Gamma$ to cool atoms below the Doppler temperature limit via 3D polarization gradient cooling (PGC), which loads atoms into ${\sim}k_B \times 0.18$-mK depth tweezers, and parity projects the number of atoms in a tweezer~\cite{schlosser_sub-poissonian_2001} to either 0 or 1. This 3D PGC is applied for 40 ms, after which we wait another 40 ms for the remaining atomic vapor from the MOT to drop and dissipate. The optical tweezer array is kept on for the entirety of the experiment.

Generating optical tweezers with an SLM results in weak out-of-plane traps that can trap sufficiently cold atoms from the MOT~\cite{singh_dual-element_2022}. This could lead to a strong background in the image or to false positives detection of single atoms, both of which affect the imaging fidelity. To avoid this issue, we apply a resonant pushout beam for 2 $\upmu$s, apply 2D PGC for 30 ms, quasi-adiabatically ramp-down the tweezer power to one-fifth of the full power, wait for 70 ms, then ramp-up the power. After this sequence, we apply 2D PGC for 180 ms with an added bias magnetic field of $0.19$ G.
Note that this sequence for removing atoms in spurious traps was not fully optimized and we believe this can be readily shortened in future work. In particular, the bias field during the 180-ms PGC segment could be more carefully optimized to reduce this time.

\subsection*{Single-atom imaging}
For single-atom imaging in the optical tweezers, we use two pairs of PGC beams in a crossed-beam configuration ($1/e^2$ diameter of 3.5 mm, 1.0 mW total). One pair is frequency detuned relative to the other pair. Each PGC beam copropagates with a repumping beam ($\sim$100 $\upmu$W) and is independently-steered. Auxiliary vertical PGC beams (not shown) aligned at a slight grazing angle along the objective axis are not used due to high background reflections off the uncoated glass cell surface. During imaging, we increase the total intensity of the 2D PGC beams by ${\sim}3\%$ and set the detuning to $\Delta = -15.5 \Gamma$ from the bare atom $\ket{6 S_{1/2}, F=4}\leftrightarrow\ket{6 P_{3/2}, F'=5}$ resonant transition. We collect scattered photons for 80 ms on a qCMOS camera (Hamamatsu ORCA-Quest C15550-20UP), which we choose for its fast readout time and its high resolution. The optical losses in the imaging system result in around 2.7\% of scattered photons entering the camera, of which 44\% are detected on the sensor due to the quantum efficiency at 852 nm. The total magnification factor of the imaging system is 5.1.

The averaged point-spread function waist radius is measured to be 1.7 pixels on the qCMOS camera, corresponding to $7.8 \ \mathrm{\upmu m}$ on the camera plane or $1.5 \, \mathrm{\upmu m}$ on the atom plane. We estimate that, accounting for a finite atomic temperature (up to $50 \ \mathrm{\upmu K}$ in this simulation) and camera sensor discretization, the ideal PSF radius should be 1.25 pixels. We leave an investigation of the discrepancy to future work.

In addition to the high-fidelity high-survival demonstrated and characterized in Fig.~2 and Ext. Data Fig.~\ref{fig:imaging-ed},\ref{fig:survival-ed}, we show in Ext. Data Fig.~\ref{fig:fidelity-fastimg-ed} imaging results acquired with an imaging time of 20 ms. Notably, this imaging data was acquired with a PGC detuning of $\Delta = -9.5\Gamma$. We measure an imaging fidelity and survival probability of 99.9571(4)\% and 99.176(1)\%.

\subsection*{Imaging model and characterization}
We now describe the binarization procedure applied to each image acquired by the qCMOS camera. For each experimental run, typically consisting of a few hundred to a few thousand iterations, we apply this procedure anew.

We identify all sites by comparing the average image with the known optical tweezer array pattern generated by the SLM. The signal for each site and each image is obtained by weighting the number of photons per pixel with a function $W(u,v)$ (Ext. Data Fig.~\ref{fig:imaging-ed}a). These weights are optimized via a quasi-Newton numerical method to maximize the imaging fidelity obtained with the model-free approach described below. This approach is agnostic of the photon distribution and relies on the consistency of the imaging outcomes. This helps guarantee that the imaging fidelity we quote is accurate, and not artificially larger due to overfitting.

We then compare the signal obtained for each site and each image with a threshold to determine if an atom has been loaded. To position the threshold and estimate the fidelity, we employ two complementary methods: an analytical model that predicts the shape of the imaging histogram by integrating the loss probability in a Poisson distribution, and a model-free approach that estimates the fidelity by identifying anomalous atom detection results in three consecutive images. The first method \emph{infers} classification errors from the shape of the photon histogram while the second method \emph{detects} errors directly; thus, the first method requires less samples to reach satisfactory accuracy. This first method is also compatible with any type of experimental runs while the second one requires to specifically acquire three consecutive images. Hence, we use the first method to position the binarization threshold in most experimental runs, as well as for site-by-site analysis; we use the second method to accurately estimate the fidelity with a single array-wide threshold. The fidelities quoted in the main text are calculated using this second method.

We first describe the analytical model that predicts of the shape of histogram, which we call ``lossy Poisson model''. We fit six parameters: the initial filling fraction (before the first image) $F$, the mean number of photons collected from the background light $\lambda_0$ and the atoms $\lambda_1$, the broadening from an ideal Poisson distribution $r_0$ and $r_1$, and the pseudo-loss probability $L$. The exact meaning of all parameters is described below.

We first derive this model in the absence of broadening from an ideal Poisson distribution. We are interested in the photon distribution given that there is no atom at a given site at the beginning of imaging $P(N=n | 0)$ and the photon distribution given that there is an atom at this site at the beginning of imaging $P(N=n|1)$, where $N$ is the number of photons collected. For the background photon distribution, we simply assume a Poisson distribution: $P(N=n|0) = e^{-\lambda_0} \lambda_0^n/n!$. For the atom photon distribution we derive an expression by considering a loss-rate model where each photon collection event (occurring with probability $\lambda_1 \mathrm{d}t$) imparts a loss probability $L/\lambda_1$. By integrating over $t\in [0,1]$ the system of equations that describes the evolution of the joint distribution of atom presence and photon count, we find the distribution given that one atom was initially present,

\begin{equation*}
\begin{split}
P&(N=n|1) = \frac{(\lambda_0 + \lambda_1-L)^n e^{-(\lambda_1+\lambda_0)}}{n!} \\
&+ \frac{L}{\lambda_1} \frac{e^{\frac{\lambda_0 L}{\lambda_1 -L}} \left(1-\frac{L}{\lambda_1} \right)^{n-1}}{(n-1)!} \\
&\times \left[\Gamma\left(n, \frac{\lambda_0}{1-L/\lambda_1}\right) - \Gamma\left(n, \lambda_1+\frac{\lambda_0}{1-L/\lambda_1}\right) \right].
\end{split}
\end{equation*}

Here, $\Gamma$ represents the upper incomplete gamma function. The real loss probability during imaging is then given by $\Tilde{L} = 1-e^{-L}$. This equation illustrates the two mechanisms that limit the imaging fidelity in experiments with single-atom imaging. The first mechanism, represented by the first term in the r.h.s. of the equation, manifests as a Gaussian/Poissonian overlap between the two peaks of the photon distribution, reflecting our ability to record a significant photon count above the imaging noise floor. Finite scattering rate, limited photon collection efficiency, background light leakage from the imaging beams or the ambient light, and readout noise from the camera contribute to this limitation. The other mechanism that limits imaging fidelity is loss of atom during imaging. This manifests as a characteristic `bridge'-like feature, and is represented by the second term in the r.h.s. of the above equation. The probability density in the bridge is small but finite across a wide range of photon counts between the two peaks of the imaging histogram~\cite{bloch_trapping_2023}.

The overall photon probability distribution is then given by $P(N=n) = F P(N=n|1) + (1-F) P(N=n|0)$. For practical purposes we empirically include a broadening of the Poisson distribution by writing $P(N=n) = F P(N=n/r_1|1) / r_1 + (1-F) P(N=n/r_0|0) / r_0$ and by effectively considering non-integer photon numbers (by replacing factorials with the gamma function). For large $n$ this amounts to considering a Gaussian distribution for either of the two peaks, but with the added benefit of including the loss through a physically-motivated derivation using a Poisson process.

In this model the true negative probability is given by $\mathcal{F}_0 = \int_0^T {P(N=n|0) \mathrm{d}n}$, where $T$ denotes the threshold; and the true positive probability, by $\mathcal{F}_1 = \int_T^\infty {P(N=n|1) \mathrm{d}n}$. Finally the imaging fidelity can be estimated as $\mathcal{F} = F \mathcal{F}_1 + (1-F) \mathcal{F}_0$ and the optimal threshold $T$ can be found by maximizing the fidelity. We find that this model performs well when predicting the shape of the histogram site-by-site (Fig.~\ref{Fig2}a), but fails when the distribution of the background or atom photons in the array is non-Gaussian.

The second method we use to characterize imaging fidelity and survival requires no assumption for the photon distribution, but considers that the imaging survival and fidelity is identical for three successive images~\cite{norcia_microscopic_2018,madjarov_entangling_2021}. We start by estimating the probability $\tilde{P}_{x_1 x_2 x_3}$ of the presence of an atom in three images being $x_1 x_2 x_3$, where $x_i$ is a Boolean, equal to 1 if there is an atom and 0 if there is none,

\begin{equation*}
\begin{split}
\tilde{P}_{111} &= S^2 F, \\
\tilde{P}_{110} &= (1-S) S F, \\
\tilde{P}_{100} &= (1-S) F, \\
\tilde{P}_{000} &= 1-F.
\end{split}
\end{equation*}
Here, $S$ is the survival probability during imaging and $F$ is the initial filling fraction. From this we can estimate the probability of \emph{detecting} $y_1 y_2 y_3$ as $P_{y_1 y_2 y_3} = \sum_{x_1 x_2 x_3} P(y_1 | x_1) P(y_2 | x_2) P(y_3 | x_3) \Tilde{P}_{x_1 x_2 x_3}$. The conditional probabilities on the detection categorization given the true atomic presence are $P(1|1) = \mathcal{F}_1$, $P(0|1) =1-\mathcal{F}_1$, $P(1|0) = 1-\mathcal{F}_0$, and $P(0|0) = \mathcal{F}_0$.

We use the method of least squares to minimize the difference between the experimental frequencies of bitstrings $y_1 y_2 y_3$ and the $P_{y_1 y_2 y_3}$ by tuning the four parameters $F$, $S$, $\mathcal{F}_0$ and $\mathcal{F}_1$. The imaging fidelity is then defined as $\mathcal{F} = F \mathcal{F}_1 + (1-F) \mathcal{F}_0$. The array-wide binarization threshold is chosen to maximize the imaging fidelity (Ext. Data Fig.~\ref{fig:imaging-ed}c). Using this method, we find an imaging fidelity $\mathcal{F} = 0.9999374(8)$, with a false positive probability $1-\mathcal{F}_0 = 7.01(8)\times 10^{-5}$ and a false negative probability $1-\mathcal{F}_1 = 5.5(1)\times 10^{-5}$; we find the survival to be $S=0.999864(2)$, slightly lower than the steady-state imaging survival probability measured by repeated imaging. Finally, we can inject the model-free survival probability into the lossy Poisson model to increase its accuracy (trying to extract the loss directly from the lossy Poisson model would indeed be inaccurate, since losses appear as a small tail feature between the two peaks of the imaging histogram). Using this approach, and fitting each site independently, we find an average imaging fidelity of 99.992(1)\%, in reasonable agreement with the model-free imaging fidelity. By setting the atom loss to zero while keeping the five other fit parameters constant for each site, we can estimate a hypothetical imaging fidelity in the absence of atomic loss of 99.999(1)\%. This analysis also illustrates that fitting the imaging histogram with a Gaussian or Poissonian model without including losses leads to overestimating imaging fidelities~\cite{levine_quantum_2021}.

Note that for data shown in this work pertaining to loading and imaging, we use images 2-4 of a set of 16,000 iterations containing each 4 successive images, since we \emph{a posteriori} realize that the survival probability and imaging fidelity are significantly higher than for images 1-3. In this latter case we measure an imaging fidelity of 0.999882(1) and survival of 0.999817(2). This could be due to remaining background vapor from the MOT loading stage, or to imperfect background atom removal during the off-plane trapped atom push-out stage. To quantify the combined survival and fidelity in each of the images, we can use the conditional probability of observing one atom given that one atom was observed in the previous image, $p(1|1)$. We find $p(1|1) =0.99963$ between the first and second images, $0.99977$ between the second and the third images, and $0.99981$ between the third and the fourth one. These numbers still can qualify as `high-fidelity and high-survival'. In principle, we could obtain the same fidelity and survival from the first image by waiting more for the background vapor to diffuse in the chamber or by extending our push-out scheme.

In the context of atomic rearrangement, we expect that several rounds of imaging and rearrangement will be required to maximize the defect-free probability, as is already common in experiments with dozens or hundreds of atoms~\cite{endres_atom-by-atom_2016,norcia_iterative_2024_mod}. Hence, the lower fidelity and survival in the first image should not impact the final efficiency of rearrangement.

\subsection*{Qubit state preparation, control, and readout}
To initialize the tweezer-trapped atoms in the $|6 S_{1/2}, F=4,m_F = 0\rangle \equiv |1\rangle$ state, we perform 5 ms of optical pumping on the 895 nm, $F=4 \leftrightarrow F'=4$ $D1$ transition. Simultaneously, we repump atoms in the $F=3$ hyperfine ground state on the 852 nm, $F=3 \leftrightarrow F'=4$ $D2$ transition. Both beams are coaligned and linearly polarized using a Glan-Thompson prism, parallel to the quantization axis defined by a 2.70-G bias magnetic field to drive $\pi$-transitions. The beams are focused to a dimension of $3.3 \,\mathrm{mm} \times 73 \,\upmu\mathrm{m}$ ($1/e^2$ waists) at the tweezer array. Angular momentum selection rules forbid the $m_F = 0 \leftrightarrow m_F' =0$ transition for $\Delta F = 0$, and the atomic population accumulates in $|1\rangle$ after multiple spontaneous emissions. We estimate a state preparation fidelity of 99.2(1)\%, inferred from the early-time contrast of the Rabi oscillations in Fig.~\ref{fig:mt-microwave}a. After preparing the atoms in $\ket{1}$,  the trap depth is adiabatically lowered to $k_B \times 55 \,\mathrm{\upmu K}$ for microwave operation.

The setup used to drive microwave transitions is described in Ext. Data Fig.~\ref{fig:uw-ed}a. Similarly to other experiments~\cite{li_toward_2009,maller_single-_2015} the RF signal from an arbitrary waveform generator (AWG, Spectrum Instrumentation M4i.6622-x8)  IQ-modulates a microwave signal generator (Stanford Research Systems SG386) set at a fixed frequency of 4.6 GHz. The signal is then frequency-doubled, filtered, passed through an isolator before being amplified to 10 W of microwave power (Qubig QDA). A 10 dBi-gain pyramidal horn emits the microwave field on the atom array at a distance of 15 cm.

For state readout we apply a resonant  $\ket{6S_{1/2},\ F=4} \leftrightarrow \ket{6P_{3/2},\ F'=5}$ pulse to push out atoms in $\ket{1}$, before imaging remaining atoms in $\ket{0}$ with the scheme described above. By measuring the off-resonantly depumped population during push-out after pumping all atoms in $\ket{F=4}$, we infer a spin-resolved push-out fidelity of 99.88(5)\%. The data in Fig.~\ref{fig:mt-microwave}, Fig.~\ref{fig:transport}, Ext. Data Fig.~\ref{fig:uw-ed}, Ext. Data Fig.~\ref{fig:site-resolved-coherence-ed}, and Ext. Data Fig.~\ref{fig:transfer-transport-ed} are not corrected for state preparation and measurement (SPAM) errors. Instead, our measurements of the coherence time and gate fidelity rely on protocols that are intrinsically insensitive to SPAM errors.

Microwave spectroscopy reveals that the initial atomic population is close to an even distribution among the $F=4$ sublevels. We measure a depumping rate of $0.064(5)\,\upmu\mathrm{s}^{-1}$ from $F=4$ to $F=3$ at our operating $D1$ optical pumping beam intensity when the $D2$ repump is shuttered off. The intensity of the $D2$ repump is increased until there is no measurable improvement in state preparation fidelity. Factors that limit the state preparation include imperfect linear polarization purity, spatial variations in the pump laser intensity due to interference fringes arising from the surface of the science glass cell, and heating incurred during the optical pumping. Modeling our magnetic field coils, we estimate that the local direction of the bias magnetic field deviates by $<10^{-5}$ radians for distances of $\sim$1 mm from the geometric center, and this has a negligible impact on the state preparation of our large scale array. Other state preparation schemes with higher fidelity have been demonstrated previously on smaller arrays and could be implemented in our system in the future~\cite{wu_sterngerlach_2019, evered_high-fidelity_2023}. 

\subsection*{Characterizing the atomic qubits}

To characterize the Rabi frequency across the array, we drive the qubit for variable times and measure the population in $\ket{1}$, both at early times (0-150 $\upmu$s) and at late times (900-1,000 $\upmu$s). We observe a spatially-varying Rabi frequency across the array (Ext. Data Fig.~\ref{fig:uw-ed}b), with a gradient that is orthogonal to the propagation axis of the microwave field, which points to a reflection off a vertical metallic optical breadboard next to the vacuum cell.

We also characterize the dephasing in the array using Ramsey interferometry. During the free-evolution time, we detune the microwave drive field by $\delta = 2\pi \times 1 \ \mathrm{kHz}$ from the average qubit frequency. The envelope of the Rabi oscillation has a Gaussian decay with a characteristic time $T_2^* = 14.0(1)$ ms. However, when considering each site individually we find an average $\langle T_2^{*(\mathrm{site})} \rangle = 25.5$ ms with a standard deviation of 3.2 ms (in the per-site case we fit the oscillation decay with the dephasing decay function from Ref.~\cite{kuhr_analysis_2005}). This shows that dephasing across the array primarily occurs because of trap depth inhomogeneities (Ext. Data Fig.~\ref{fig:loading-ed}d): assuming a Gaussian distribution of trap depth with a standard deviation $\delta U$, the qubit frequencies in the array also follow a Gaussian distribution, which results in an ensemble-wide dephasing time $T_2^{* \mathrm{(inh)}} = \sqrt{2} \hbar/(\eta \, \delta U)$ where $\eta$ is the ratio of the scalar differential polarizability of the hyperfine ground states to their polarizability at the fine structure level~\cite{kuhr_analysis_2005}. On the other hand, finite atomic temperature limits the per-site dephasing time $T_{2}^{* (\mathrm{site})}$. We observe an uneven distribution of $T_2^*$ across the atom array (Ext. Data Fig.~\ref{fig:site-resolved-coherence-ed}b), with a significantly lower $T_2^*$ measured for atoms trapped in tweezers at 1055 nm than for those trapped in the bottom half of tweezers at 1061 nm. This discrepancy could be due to worse optical aberrations in these areas that decrease the efficiency of polarization-gradient cooling, or due to different intensity noise profiles from the different fiber amplifiers or SLMs used on the two pathways. These data reveal that further investigation of noise sources specific to lasers or tweezer pathways could elucidate limiting factors on coherence times in neutral atom arrays.

In order to relate $T_2^*$ and trap depth inhomogeneity or atomic temperature, the parameter $\eta$ can be calculated as the ratio of the differential light shift of the hyperfine states to the electronic ground state light shift, which yields $\eta = 1.50 \times 10^{-4}$. (At the few-percent accuracy level, it becomes important to account for higher-order processes~\cite{rosenbusch_ac_2009,carr_doubly_2016}, but such accuracy is not required here). We corroborate this value by experimentally measuring the differential light shift via Ramsey interferometry at different depths (Ext. Data Fig.~\ref{fig:uw-ed}c). We find $\eta = 1.3(1) \times 10^{-4}$, in reasonable agreement with the theoretical value. This allows us to estimate the atomic temperature during microwave operation as~\cite{kuhr_analysis_2005} $T = \sqrt{e^{2/3}-1}\times 2\hbar/(\eta k_B \langle T_2^{*(\mathrm{site})} \rangle) \approx 4.3\ \mathrm{\upmu K}$ (assuming the temperature is sufficiently homogeneous to invert the fraction and the mean). This temperature may differ from the effective atomic temperature during other points of the experimental sequence that do not include the ramp-down and state preparation steps that may decrease and increase the temperature respectively.

\subsection*{Dynamical decoupling}

In order to extend the operation time of a realistic quantum processor well beyond the dephasing time of the array, we can apply dynamical decoupling on the atomic qubits. We empirically find that a period of 12.5 ms yields the longest dephasing time for the reduced trap depth of $k_B \times 55 \, \upmu$K.

We vary the number of symmetric XY16 cycles and we obtain the coherence contrast by applying a final $\pi/2$ pulse with phase 0 or $\pi$. Subtracting the population difference in these two cases yields the coherence contrast after the dynamical decoupling sequences.

We investigate in Ext. Data Fig.~\ref{fig:uw-ed}g the coherence time $T_2$ as a function of the trap depth for two different periods between $\pi$ pulses (only for atoms trapped with the fiber amplifier at 1061 nm), 12.5 ms and 6.2 ms. We attribute the different optimal periods at different depths to a trade-off between the unfiltered noise at a specific dynamical decoupling period~\cite{viola_dynamical_1999} and the effective depolarization induced by each $\pi$ pulse. At the full trap depth, we measure a coherence time of 3.19(5) s, which still constitutes a record for hyperfine qubits in a tweezer array.

Considering the Raman scattering rate at a trap depth of $k_B \times 0.18$ mK, we expect that a significantly longer coherence time should be achievable. Based on this observation and the discrepancy in coherence time between atoms trapped at 1061 nm and 1055 nm seen in site-resolved data (Ext. Data Fig.~\ref{fig:site-resolved-coherence-ed}c), we posit that the observed coherence time is limited by intensity noise due to the trapping lasers or the SLMs. We leave further investigation to future work.

\subsection*{Single-qubit gate randomized benchmarking}

We measure our single-qubit gate fidelity via randomized benchmarking, similarly to Refs.~\cite{xia_randomized_2015,nikolov_randomized_2023}. For each given length $n$, we select $U_{n-1}, \dots, U_0$ at random from the 24 unitaries composing the Clifford group. We then apply $U_{-1} U_{n-1} \cdots U_0$ where $U_{-1}$ is the inverse of $U_{n-1} \cdots U_0$. We decompose Clifford gates into elementary rotations around Bloch sphere axes using the $zyz$ Euler angles. Rotations around $z$ are implemented by offsetting the phase of all following $x$ and $y$ rotations~\cite{mckay_efficient_2017}.

Due to the inhomogeneous Rabi frequency, each rotation must be applied using length error-resilient composite pulses. Among common families of error-resilient pulses~\cite{wimperis_broadband_1994,cummins_tackling_2003,kukita_short_2022}, we find that SCROFULOUS performs the best in our case. The SCROFULOUS implements a rotation of angle $\theta$ around the axis indexed by the angle $\phi$ on the Bloch sphere equatorial plane (abbreviated as $\theta_\phi$) with a symmetric composite pulse $(\theta_1)_{\phi_1} (\theta_2)_{\phi_2} (\theta_3)_{\phi_3}$ where $\theta_1 = \theta_3 = \mathrm{arcsinc}(2 \cos(\theta/2)/\pi)$, $\phi_1 = \phi_3 = \phi + \arccos{\left(-\frac{\pi \cos{\theta_1}}{2\theta_1 \sin(\theta/2)}\right)}$, $\theta_2 = \pi$ and $\phi_2 = \phi_1 - \arccos{\left(-\frac{\pi}{2\theta_1} \right)}$. In our implementation, the average pulse area for a random Clifford unitary is $2.02\pi$.

We fit the decay of the final population with the number of applied Clifford gates as $\frac{1}{2} + \frac{1}{2}(1-d_0)(1-d)^n$ where $d_0$ stems from SPAM errors, $d$ is the average depolarization probability at each gate and $n$ is the number of gates. The average Clifford gate fidelity is then given by~\cite{knill_randomized_2008}: $F_c = 1-d/2$.

Even though the measured single-qubit gate fidelity is competitive with other state-of-the-art atom arrays experiments~\cite{wang_single-qubit_2016,graham_multi-qubit_2022,ma_universal_2022,bluvstein_logical_2024}, single-qubit gate fidelities ${>}0.9999$ have been reported~\cite{sheng_high-fidelity_2018,nikolov_randomized_2023} in smaller arrays. Moreover, the maximal theoretical fidelity achievable for a given dephasing time is~\cite{xia_randomized_2015} $\mathcal{F} = \frac{3}{4} + \frac{1}{4(1+0.95(t/T_2^*)^2)^{3/2}}$ where $t$ is the average time needed to apply a Clifford gate, $t = \langle \theta \rangle / \Omega$; $\langle \theta \rangle$ being the average pulse area per Clifford gate. Hence, gate fidelities higher than 0.99999 should be achievable solely based on this value.

Beyond infidelities due to decoherence, other parameters that may limit single-qubit gate fidelities are: (a) amplitude errors due to instabilities in the microwave power; (b) phase errors due to the microwave setup; (c) phase errors due to optical tweezer intensity noise; (d) phase errors due to magnetic field noise. We are interested in which of these factors is limiting the gate fidelity. We rule out (a) because we observe that the Rabi frequency is very stable shot-to-shot (variations of less than 0.1 \%), and we estimate that such variations should be completely suppressed by the SCROFULOUS pulse. We also rule out (c) since reducing the trap depth further does not significantly improve the randomized benchmarking results (Ext. Data Fig.~\ref{fig:uw-ed}g), and the fidelity is identical for atoms trapped in tweezers at 1055 nm and 1061 nm (unlike $T_2^*$ and $T_2$). Although we cannot formally rule out (b), we estimate that it is unlikely since active components in the microwave setup have a very low phase noise, and we observe a sub-10 Hz linewidth of the microwave signal with a spectrum analyzer.

We also notice a dominant phase noise at 60 Hz in the qubit array due to the mains AC voltage. We measure the intensity of this noise with a spin-echo sequence, where the time between each pulse is $\tau = 1/(2\times 60 \ \mathrm{Hz})$ (Ext. Data Fig.~\ref{fig:uw-ed}e). Although this low-frequency noise cannot by itself explain the single-qubit gate fidelity loss, it points out more generally to residual magnetic field noise that could be mitigated by shielding the vacuum cell, upgrading the current sources driving the magnetic field coils, and/or by operating at MHz-scale via Raman transitions. This can be achieved, for instance, by utilizing the amplitude-modulation setup used for Raman sideband spectroscopy.

\vspace{0.2cm}

\subsection*{Raman sideband spectroscopy with amplitude-modulation setup}

In order to measure the axial and radial trapping frequencies we use a Raman setup based on amplitude modulation of a laser beam\cite{levine_dispersive_2022}. The laser beam, red-detuned by 345 GHz from the $D1$ electronic transition in $^{133}$Cs, is phase-modulated using an resonant electro-optic modulator at 9.2 GHz (Qubig) before reflecting twice off a highly dispersive chirped Bragg grating (Optigrate CBG-894-90) that transforms phase modulation into amplitude modulation. Two amplitude-modulated beams with different wavevectors $\boldsymbol{k_1}$ and $\boldsymbol{k_2}$ drive sideband transitions, akin to previous works with mode-locked lasers used to address the motion of trapped ions~\cite{hayes_entanglement_2010,inlek_quantum_2014}. A schematic of the setup is shown in Ext. Data Fig.~\ref{fig:raman-sideband-ed}a.

In this configuration, the effective Lamb-Dicke parameter is $\eta_{\alpha}^{\mathrm{LD}} = |(\boldsymbol{k_1}-\boldsymbol{k_2})\cdot \boldsymbol{\alpha}| \sqrt{\frac{\hbar}{2 m \omega_{\alpha}}}$, where $m$ represents the mass of cesium-133, and $\alpha$ denotes the radial or axial motion (with unit vector $\boldsymbol{\alpha}$). Out of 1 W of fiber-coupled amplitude-modulated laser light, each beam has 1-5 mW of laser power and a Gaussian $1/e^2$ diameter of ${\sim}2$ mm. The sideband spectroscopy results are shown in Ext. Data Fig.~\ref{fig:raman-sideband-ed}b,c, with radial and axial trapping frequencies measured to be, respectively, 29.30(4) kHz and 5.64(3) kHz. From this measurement we infer a $1/e^2$ tweezer waist $w_0 = 1.17(6) \ \mathrm{\upmu m}$. From the lineshape fit, we extract standard deviations across the array of 4.7 kHz and 1.9 kHz, respectively. Note that this measurement was done with atoms in the 1061-nm tweezer array.

\subsection*{Atom transport}
We create 10 transport tweezers using 1055 nm light through two AODs (Gooch \& Housego AODF 4085), mounted in a crossed configuration and with an active aperture of ${\sim}15$ mm diameter. We map the output after the pair of AODs to the back aperture of the objective using a telescope with 3:2 demagnification to match the same beam size at the back aperture of the objective as the beam from the SLM trapping tweezers. 

The 1055-nm light for transport is split from the same laser source that makes tweezers in the center of the array (see Ext. Data Fig.~\ref{fig:loading-ed}a). The 1055-nm static and transport tweezers are then recombined with polarization, and combined with 1061-nm light with a polarizing beam splitting cube as well. These two pathways are not used concurrently for the long-distance coherent transport demonstration in Fig.~\ref{fig:transport} or in Ext. Data Fig.~\ref{fig:transfer-transport-ed}b,c, and d. We plan to switch in the near term to combining the 1055-nm and 1061-nm light using a dichroic mirror, such that we can use the power in the 1055-nm path for both static and transport tweezers simultaneously without loss.

For the atomic movement, we use an adiabatic sine trajectory described by $x = \frac{1}{\pi}\sin(\pi t) + t \quad (t,x\in [-1,1])$. We find that we can execute a single move faster with the constant jerk trajectory~\cite{bluvstein_quantum_2022} (which we use for Fig.~\ref{fig:transport}b and SI Fig.~4), but that the adiabatic sine trajectory incurs less heating: in the harmonic oscillator approximation, the increase in the average radial motional quanta $\Delta N$ for an adiabatic sine trajectory scales as $\Delta N \propto \frac{D^2}{\omega^5 T^6}$ where $D$ is the distance of the trajectory, $T$ is the time of the trajectory, and $\omega$ is the trap frequency. In the case of a constant jerk trajectory, $\Delta N \propto \frac{D^2}{\omega^3 T^4}$.

Note that in the coherent transport data, the tweezer depth change along the trajectory is compensated with RF power which we calibrate beforehand with static tweezers at each position. We believe the transport fidelity can be further increased with more careful compensation of the trap depth including the AOD lensing effect in the future.

\vspace{0.2cm}
\subsection*{Randomized benchmarking of coherent transport}

Coherent transport is achieved by suppressing dephasing during transport with dynamical decoupling. By evaluating the coherence contrast after 80 moves, we empirically find that the asymmetric XY4 sequence~\cite{souza_robust_2012} performs best (implemented using bare pulses).
To perform interleaved randomized benchmarking~\cite{magesan_efficient_2012}, we fix a total number of single-qubit gates $N$ drawn from the Clifford group $C_1$. We then interleave $M (< N)$ total moves between the first $M$ gates (atoms are held for $\sim$54 $\upmu$s between moves), after which we apply the remaining $N-M$ gates to keep the total number of gates $N$ constant and then apply the inverse of these gates. For the return probability data shown in Fig.~\ref{fig:transport} and Ext. Data Fig.~\ref{fig:transfer-transport-ed}a-d, we average over 72 sequences of random gates for each number of moves and apply $N=80$ total random single-qubit Clifford gates. For the static and transported return probabilities, we apply the same single-qubit control sequence, including the XY4 dynamical decoupling. As in the case of randomized benchmarking, we utilize SCROFULOUS pulses for implementing the Clifford gates.

During each move of the benchmarking sequence, we apply XY4 in a \emph{transformed Clifford frame}. Previous works have examined the interplay of dynamical decoupling and quantum operations by; e.g., studying a system Hamiltonian in the ``toggling frame'' induced by dynamical decoupling pulses~\cite{morong_engineering_2023}. Here, we use related ideas but examine the decoupling operations in the frame rotated by the previously applied Clifford gates.

For instance, ignoring the Clifford gates between moves $k-1$ and $k$, it is possible to concatenate two XY4 sequences $X-Y-X-Y$ (with a symmetry operation) to obtain an XY8 sequence $X-Y-X-Y-Y-X-Y-X$ that yields higher-order dephasing (and pulse-length error) suppression. However, the random Clifford gate $U_k$ between the two sequences will cancel this effect by twirling the second XY4 sequence with respect to the first one. Thus, we can ``counter-twirl'' the second XY4 sequence by applying it in a specific Clifford frame: the Pauli operator $P$ becomes $P' = U_k^{\dagger} P U_k$. Up to a global phase, $U_k^{\dagger} X U_k$ and $U_k^{\dagger} Y U_k$ are two distinct elements of $\{X,Y,Z\}$, because $U_k$ is a Clifford gate. If one of these two unitaries is $Z$, we further conjugate with a Hadamard gate $H$ (or the equivalent basis change unitary between $Y$ and $Z$) to map these two unitaries into $X$ and $Y$, or $Y$ and $X$. This can easily be generalized to $n$-qubit Clifford gates. A paradigmatic example is the transport between the storage and interaction zone to apply a $\mathcal{C\! Z}$ gate: since $\mathcal{C\! Z} (X\otimes X) \mathcal{C\! Z} = -Y\otimes Y$, we can appropriately transform the decoupling sequence applied during the return move. This could also be extended to yield higher-order sequences, such as XY16. Importantly, typical architectures for fault-tolerant quantum computation (FTQC) use almost exclusively Clifford gates~\cite{campbell_roads_2017} (e.g., past the initial generation of noisy magic-state inputs, all gates are Clifford). Therefore this technique is fully applicable to FTQC.

At the end of the randomized benchmarking sequence, we measure both the atomic survival and the return probability (note that we apply a final $\pi$ pulse to map the return state to the non-pushed-out state $\ket{0}$). We fit the atomic survival to a clipped Boltzmann distribution $S_n = 1-\exp(-1/(a+bn))$ where $a$ and $b$ are respectively the normalized initial temperature and normalized temperature accumulated per move. For the selected durations for interleaved randomized benchmarking, we find that $a$ is negligible. We then fit the return probability to $(1-\exp{(-1/bn)}) \cdot \left(\frac{1}{2}+\frac{1}{2} (1-d_0')(1-d')^n \right)$, where $d'$ is the depolarizing probability for coherence, not accounting for atom loss. Owing to the randomized benchmarking procedure, coherence loss also includes the impact of XY4 dynamical decoupling since it would not be necessary without transport. We then extract the instantaneous fidelity after $n$ moves as $F_n = \left(1-\frac{d'}{2} \right)\frac{1-\exp{(-1/b(n+1))}}{1-\exp{(-1/bn)}}$. Note that this is the most conservative approach and amounts to considering that the channel infidelity due to losses is equal to the loss probability itself. In the context of fault-tolerant quantum computation, losses could be directly detected, leading to a higher tolerance to such errors than to Pauli errors. One could therefore assimilate loss to a depolarizing channel, which would significantly increase late-time instantaneous fidelities in Fig.~\ref{fig:transport}, Fig.~\ref{fig:transfer}, and Ext. Data Fig.~\ref{fig:transfer-transport-ed}. It is worth noting that losses are subdominant for early-time IRB results presented in Fig.~\ref{fig:transport} and Ext. Data Fig.~\ref{fig:transfer-transport-ed}, therefore the quoted early-time fidelity in these figures is independent of the specific model we use for losses. To compute the 68\% confidence interval we bootstrap $b$ and $d'$ using the fit results and covariance matrix. We corroborate the obtained fidelity with a simple exponential fit for the first few data points of the return probability, where losses are negligible, and find similar early-time fidelities and error bars. We also notice that the shorter move of 270 $\upmu$m has a slightly lower early-time instantaneous fidelity of $99.935(2)\%$ compared with with the 610 $\upmu$m-move ($99.953(2)\%$). We believe the discrepancy is likely due a trap depth calibration imperfection, and leave further investigation to future work.

For some applications, one might desire to optimize on the speed of movement and use a deeper trap to do so. In Ext. Data Fig.~\ref{fig:transfer-transport-ed}d, we show that atoms can be moved by 270 $\upmu$m in 400 $\upmu$s with a trap depth of $k_B \times0.92$ mK, at the cost of a reduced fidelity of $\sim$99.85\%. Comparing Ext. Data Fig.~\ref{fig:transfer-transport-ed}b and Ext. Data Fig.~\ref{fig:transfer-transport-ed}d illustrates a trade-off pertaining to coherent transport: while atoms can be moved faster by increasing the trap depth $U$, the associated transport fidelity for small number of moves is also reduced. In the limit where noise is entirely induced by tweezer intensity fluctuations, this can be understood by noticing that the dephasing strength scales as $U$ when the required duration for long-distance transport merely scales as $U^{-1/2}$. We note that experimentally in static traps, we find an even stronger scaling of coherence time than linear in $U$, likely due to other sources of noise (Ext. Data Fig.~7g).

\subsection*{Atom transfer between SLM and AOD tweezers}

To transfer atoms between static and dynamic traps, we generate an evenly-spaced grid of $15 \times 14$ AOD tweezers (with a spacing five times that of SLM sites, Fig.~\ref{fig:transfer}a), of which 195 sites overlap with SLM tweezers generated with the 1061-nm tweezer laser (out of 11,397 sites in the SLM array). The focal planes are matched by imprinting a Zernike defocus polynomial using the SLM. The position of each SLM site is adjusted in the WGS algorithm to match the corresponding AOD site, first by matching the point-spread function on the qCMOS camera, and then by optimizing the transfer survival. For the data shown in Fig.~\ref{fig:transfer}, the SLM trap depth is ramped down to ${\sim}k_B \times 0.14$ mK, which we find is optimal for transfer into $k_B \times0.28$ mK-deep AOD tweezers. We note that adiabatic ramping between full depth and this depth does not incur noticeable losses.

For hand-optimized trajectories shown in Fig.~\ref{fig:transfer}b, the AOD trap depth is quadratically increased over the course of 48\% of the total ramp-and-move duration, after which the AOD trap is moved with a constant jerk trajectory by 2.4 $\upmu$m during the remaining 52\%. These ratios, as well as the ramp and trajectory used, are set to empirically maximize atom survival.

As an alternative, we propose and implement a machine-learned procedure for faster (or equivalently, higher-survival) atom transfer, for which the AOD trap depth and position can be simultaneously changed (Ext. Data Fig.~\ref{fig:transfer}c). For both trap depth and position, 14 points---from which ramps are obtained by cubic interpolation---are adjusted by a machine learner~\cite{wigley_fast_2016} for a fixed one-way duration of 400 $\upmu$s and 60 consecutive one-way transfers. This trajectory is inverted to merge and drop off atoms back into static traps. 

For data shown in Fig.~\ref{fig:transfer}, AOD tweezers are repeatedly ramped up and moved away from the corresponding SLM sites by 2.4 $\upmu$m, and then held static for 100 $\upmu$s. The direction of motion is as pictured in Fig.~\ref{fig:transfer}a, and does not match the direction of transport used in Fig.~\ref{fig:transport} because cylindrical lensing is not detrimental at the speeds being reached. During the 100 $\upmu$s wait time, SLM tweezers are turned off after which they are turned back on, such that atoms held in SLM tweezers and not AOD tweezers are dropped. This enables us to ensure that atoms that may have remained in traps rather than being successfully picked up, are not counted towards survival. At the end of the sequence atoms are imaged again in SLM tweezers. We use the same dynamical decoupling sequence as for the AOD-only transport experiment, including notably the transformed Clifford frame technique. Unlike for long-distance transport, we find that the survival as a function of the number of transfers has an exponential component---likely due to experimental imperfections. Hence, we fit it to: $S_n = p_0 p^n (1-e^{-1/bn})$. In order to accurately distinguish between depolarizing effects and atom loss in the IRB return signal, we fit the return probability as $R_n = S_n D_n$ where $D_n = p_0' p'^n (1-e^{-1/b' n})$. The fidelity per move is then extracted as $F_n = (1/2+S_{n+1}/2S_n) \, D_{n+1}/D_n$. The uncertainty is obtained by first bootstrapping fitting parameters for $S_n$, and then, for each sample, by bootstrapping those for $D_n$. Unlike for AOD-only transport, the choice of convention used to account for losses impacts the early-time instantaneous fidelity. In a scenario where losses can be directly detected, one could assign an infidelity from loss equal to half the loss probability---as in a depolarizing model. In this scenario, the instantaneous fidelity quoted in Fig.~\ref{fig:transfer}d would increase from 99.81(3)\% to 99.88(3)\%.

\subsection*{Combined atom transfer and move}
In order to combine atom transfer and long moves, we change the static tweezer configuration to one featuring alternating spacing as shown in Ext. Data Fig.~\ref{fig:transfer-transport-ed}e. This configuration is motivated by the compatibility with diagonal motion (as schematized in SI Fig.~5), and by the observation of additional losses in the absence of wider pathways for transport.  We did not attempt to optimize SLM parameters in the original configuration to mitigate these losses and leave further investigation to future work. We include a simulation for the out-of-plane interference for this spatial distribution of tweezers now in Ext. Data Fig.~\ref{fig:out-of-plane-ed}. Apart from a slightly lower imaging fidelity---which, in the context of fault-tolerant quantum computation, matters much less than, e.g., for quantum simulation---we do not expect this array to exhibit different metrics from the configuration characterized in the rest of this paper. We then transport atoms with $8\times 6$ AOD tweezers (of which 47 are overlapped with one of the 11,416-site 1061-nm tweezer array) spanning $285 \ \upmu\mathrm{m} \times 204 \ \upmu \mathrm{m}$.

The combined transfer and move sequence is realized as follows (Ext. Data Fig.~\ref{fig:transfer-transport-ed}e): we first apply a single-qubit gate, pick up atoms from the highlighted sites on the top side of the array, and then perform a constant jerk movement for the initial separation move. We then implement the $375 \ \upmu$m-move using an adiabatic sine trajectory, apply a single-qubit gate while atoms are held in AOD tweezers on the bottom side of the array (shown in highlighted locations), before applying the reverse move and transfers. Timescales for each operation are shown in Ext. Data Fig.~\ref{fig:transfer-transport-ed}e. During the pick-up operation, AOD tweezers are ramped up from 0 to $k_B \times 0.28$ mK while SLM tweezers are ramped down from $k_B \times 0.18$ mK to $k_B \times 0.06$ mK (the trap depth used for the measurement of coherence times in Fig.~\ref{fig:mt-microwave}). Possible deleterious effects from the repeated ramps on static atoms are captured by the `static' data in Ext. Data Fig.~\ref{fig:transfer}f: the equivalent idle fidelity is $>99.96\%$.

We notice no significant exponential component in the survival signal. Therefore, when we evaluate the instantaneous fidelity using the technique described in the ``Randomized benchmarking of coherent transport'' section, the early-time estimate of instantaneous fidelity is not impacted by the choice of convention for handling loss. We anticipate that the timescales used here can be considerably sped up by leveraging machine-learning to optimize various trajectories and ramps, as demonstrated in Fig.~\ref{fig:transfer}. Additionally, we envision integrating the short move to split (and merge) AOD and SLM tweezers with the longer move in a single, curved trajectory.

\FloatBarrier
\putbib
\end{bibunit}

\end{document}


\title{Supplementary Information: Outlook to Computing}
\maketitle


We present in this Supplementary Information a comprehensive outlook on the ingredients necessary for implementing a universal quantum computer utilizing the platform in the main text of ``A tweezer array with 6100 highly coherent atomic qubits'', focusing on prerequisites that represent a challenge due to scale. For instance, we leave aside the question of mid-circuit readout in alkali atoms, for which there is an important ongoing research effort\cite{shea_submillisecond_2020, deist_mid-circuit_2022,graham_multi-qubit_2022}, since there is no clear field-of-view or power limitation due to the large size of the array.

While a digital quantum computer is only one of the applications of this atom array, it is the one that requires the highest degree of control, and thus the strategies laid out in this section can be also utilized to implement, e.g., an analog quantum simulator with some degree of local control. 

To this end, after first presenting an overview of a quantum processing zone-architecture compatible with our qubit array (SI §\ref{Zone Architecture}), we discuss in depth data-based estimates for rearrangement timing (SI §\ref{Rearrangment}), coherent control at scale (SI §\ref{Coherent transport}), and plans for implementing two-qubit gates (SI §\ref{Rydberg Gates}). Notably, based on long-distance transport data compatible with rearrangement moves, experimentally implemented algorithms, and timed processes, we conclude that the array can be rearranged quadrant-wise in less than 140 ms in parallel or in less than 525 ms sequentially. 

This investigation into the usability of the array for quantum computing yields insights into challenges to be expected with scaling up the array size, ultimately revealing a feasible near-term pathway to large scale universal quantum computation with $\sim$6,000 qubits.

\begin{figure*}[hbt!]
	\centering
	\includegraphics[width=80mm]{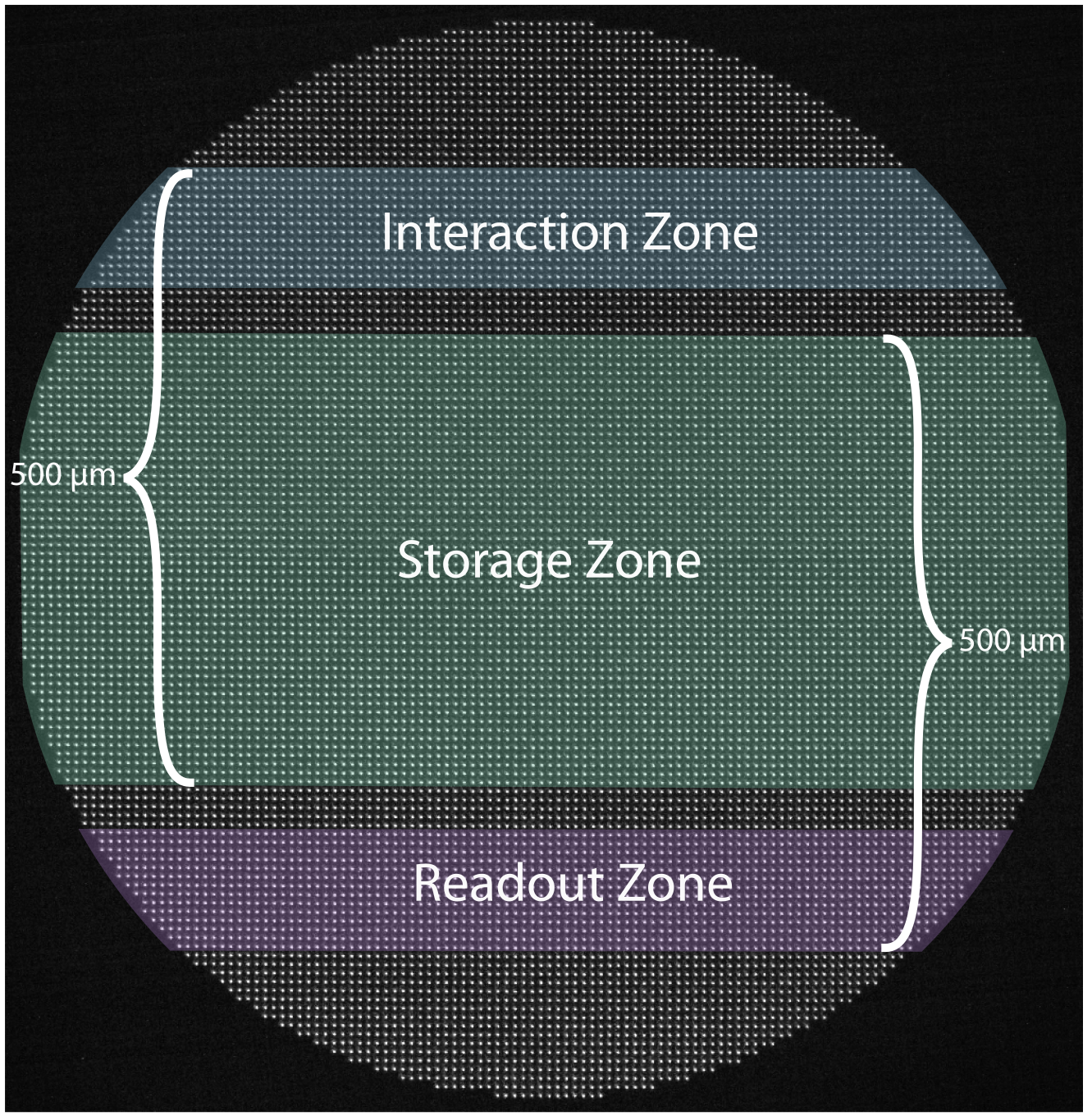}
	\caption{\textbf{Possible layout for a large-scale zone-based atom array quantum processor.} A proposed configuration for a zone-based architecture, shown on the 12,000-site tweezer array for scale. The storage zone contains enough sites to hold over 6,100 atoms, and the interaction and readout zones are within 500 $\upmu$m of any atom in the storage zone, within the field of view of the AODs we use for movement.}
 \label{fig:zone-architecture}
 \end{figure*}

\section{Zone Architecture overview}
\label{Zone Architecture}
To execute quantum operations, we envisage the usage of a zone structure~\cite{kielpinski_architecture_2002, bluvstein_quantum_2022} for entanglement and readout in combination with global and local hyperfine control. There are multiple constraints that must be considered to ensure that all operations can be realized without fidelity loss and with reasonable timescales. We briefly review these constraints in this section, and carefully analyze them in the rest of the SI. We plan to begin an experiment by rearranging (see the following section for details) atoms into a storage zone as sketched out in SI Fig. \ref{fig:zone-architecture}, reproduced from Ext. Data Fig.~10a in the main text. This zone is capable of holding over 6,100 atoms with the current spacing. One could additionally imagine packing qubits more closely in this zone with tighter tweezer spacing (using different wavelengths as simulated in Ext. Data Fig.~3c). A diagonal configuration, as we use for a proof-of-principle characterization of pick-up and transfer across $375\mu$m with 47 dynamic tweezers in a large scale static array (Ext. Data Fig.~10e,f), could also be explored.

The interaction and readout zones are within reach of the storage zone utilizing a set of Gooch \& Housego crossed acousto-optic deflectors (G\&H AODF 4085) that have an angular field of view of ${\sim}45$ mrad and an active aperture diameter of 15 mm, resulting in an effective field of view of $>500 \ \upmu\mathrm{m}$ after 3:2 demagnification. We have demonstrated movement of over $500 \ \upmu\mathrm{m}$ using these AODs, presented in Fig. 5a of the main text and in SI Fig.~\ref{fig:transport-time}.

As we demonstrate in the main text, and present in Fig.~5d, the movement distance of ${\sim}600$ $\upmu \mathrm{m}$ can be traversed while maintaining qubit coherence during several moves. In order for the AODs to be accessible to every part of the array, we ultimately plan to implement an approach based on four pairs of AODs whose accessible regions form overlapping quadrants. This would allow for parallel movement during both rearrangement and quantum circuits, as well as for every part of the array to have access to any other.
Alternatively, we could add a fast-steering mirror in one of the Fourier planes of the 4f mapping of the AOD tweezers to the atomic plane, and use this mirror to switch between different sections of the array so that the AODs can be used to move atoms. A commercially available option for such a fast-steering mirror is the Thorlabs FSM75, which has a specified 5 ms switching time and position repeatability of better than 20 nm in the atom plane.

The interaction and readout zones are each separated from the storage zone by ${\sim}40$ $\upmu \mathrm{m}$, and we plan to shape the Rydberg and imaging beams that form these zones into flat-top beams utilizing SLMs. We envision additionally applying a relative light-shift between readout and other zones in order to prevent scattered photons from causing other qubits to decohere~\cite{urech_narrow-line_2022, norcia_midcircuit_2023, hu_site-selective_2024}. The readout zone, as designated in SI Fig.~\ref{fig:zone-architecture}, would have over 1,400 SLM-generated sites, and for the gate spacing necessary for two-qubit Rydberg gates in the interaction zone, we would be able to perform around 500 two-qubit entangling gates in parallel.

\begin{figure*}[hbt!]
	\centering
	\includegraphics[width=80mm]{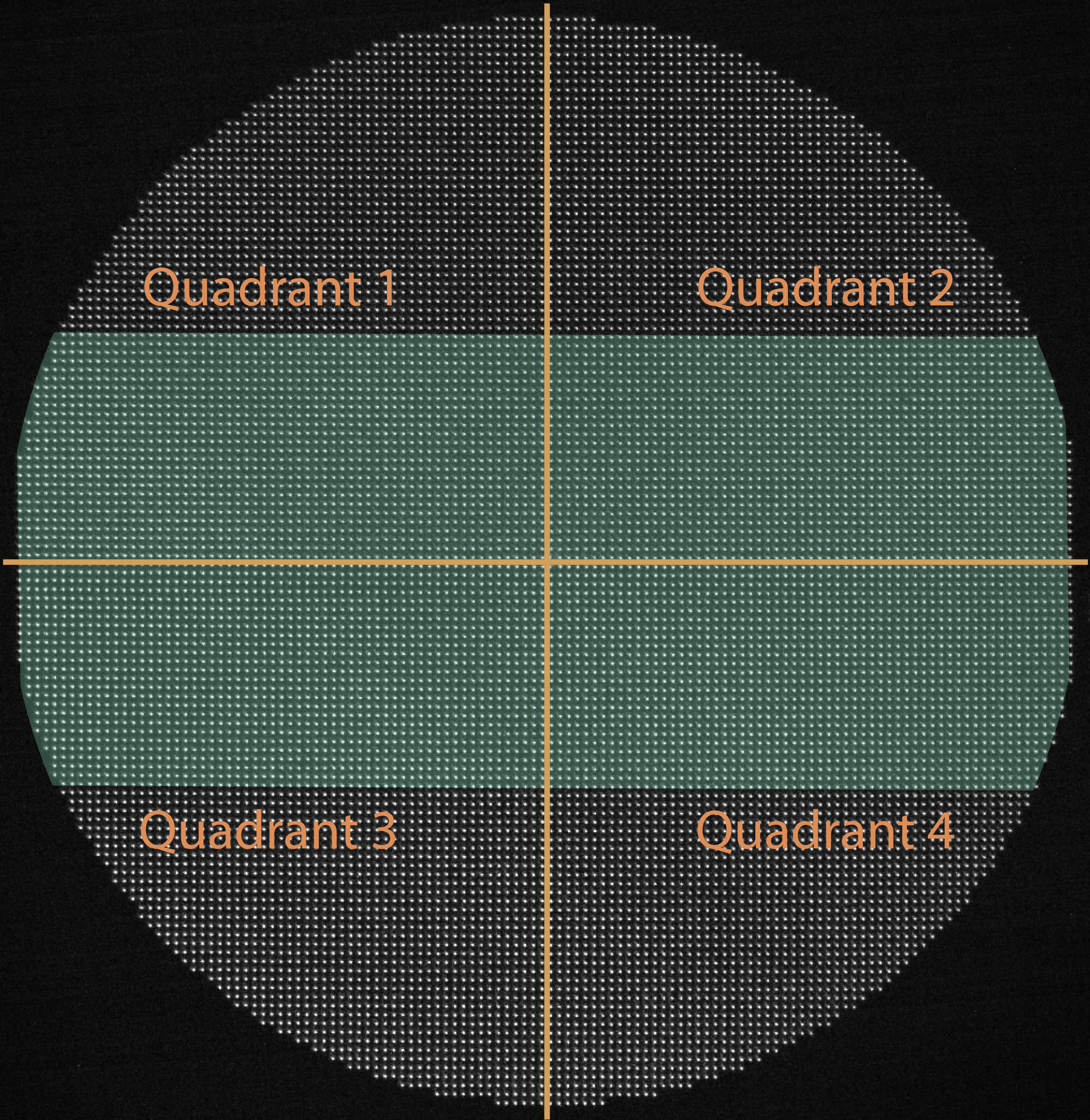}
	\caption{\textbf{Quadrant-based rearrangement.} A sketch of the quadrants that we propose to rearrange using large-aperture AODs. Combining pairs of AODs that cover each region will allow for parallel rearrangement. One could also rearrange quadrants in succession one after another, utilizing one set of AODs and a fast-steering mirror for switching.
 }
 \label{fig:four-quadrant}
\end{figure*}
\section{Rearrangement}
\label{Rearrangment}
Atom rearrangement using optical tweezers has so far been demonstrated for target array sizes of up to a few hundred to approximately a thousand sites~\cite{gyger_continuous_2024,norcia_iterative_2024_mod,pause_supercharged_2024}. Rearranging into a large-scale array with thousands of sites has not yet been achieved and remains an outstanding challenge. As discussed in the main text and previous sections, the field of view of commercially available AODs are currently insufficient to cover the entirety of our array size. To address this limitation, we plan to divide the 12,000-site array into four quadrants, as shown in SI Fig.~\ref{fig:four-quadrant}, and perform rearrangement within each quadrant. This can be achieved either by using multiple AOD pairs or, alternatively, a pair of AODs combined with a fast-steering mirror, as discussed above. 

In this section, we estimate the time required for rearrangement based on our experimental data and simulations, demonstrating that it is both feasible and achievable within a practical timeframe.

\subsection{Tetris rearrangement for each quadrant}

\begin{figure*}[t!]
	\centering
	\includegraphics[width=90mm]{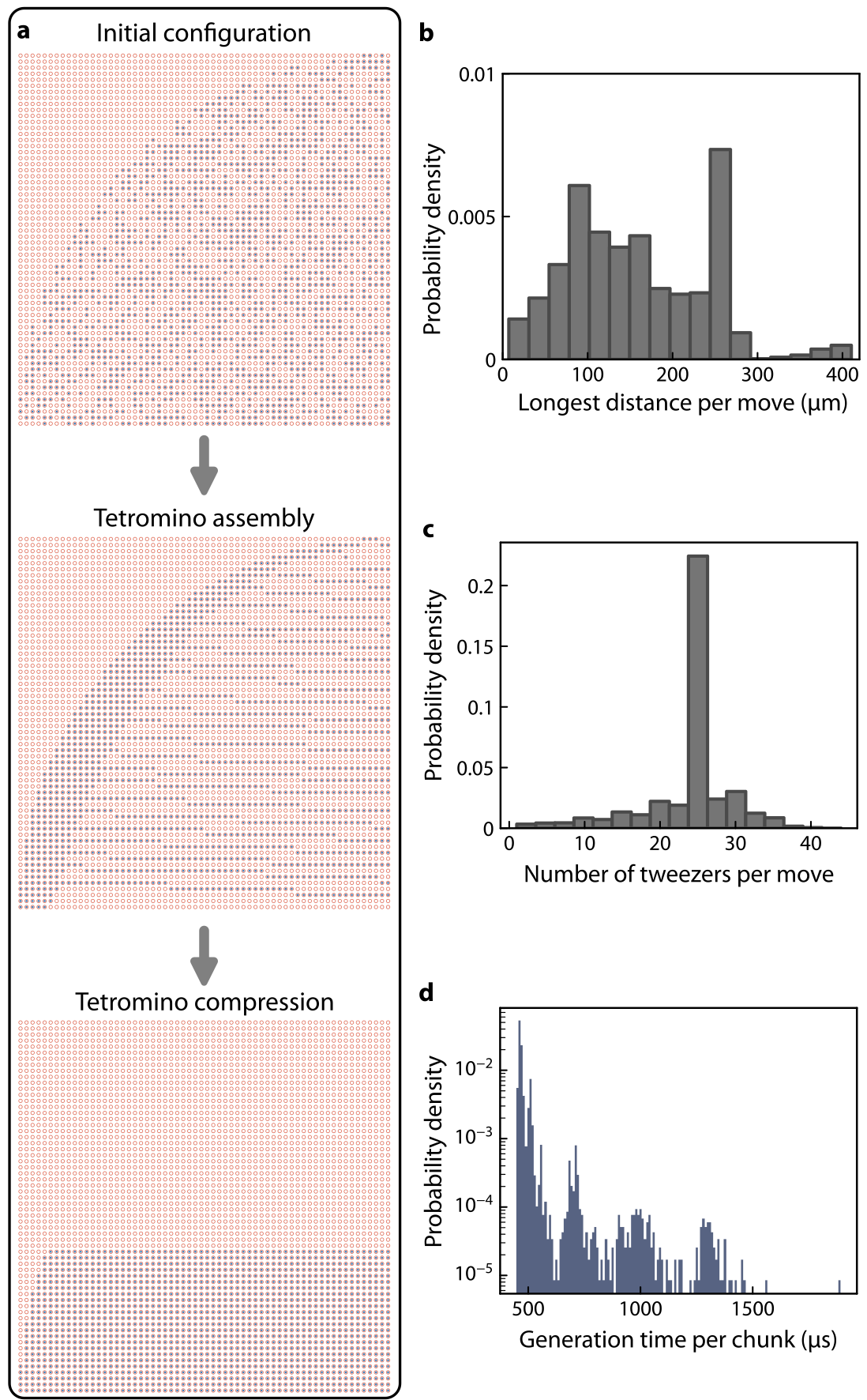}
	\caption{\textbf{Tetris sorting algorithm adaptation for a quadrant.} \textbf{a,} Illustration of the algorithm for a randomly selected example. \textbf{b,} Histogram of the longest distance in each row-based rearrangement move (usually composed of several atom trajectories), averaged over $\sim$1,000 iterations of random initial configurations. This longest distance determines the move's total duration. \textbf{c,} Histogram of the number of atoms that are moved in parallel during a move. \textbf{d,} Histogram of the required computation time for each waveform chunk of 524k samples for our parallel waveform generation algorithm (with 64 parallel movements being summed). The average generation time is 488 $\upmu$s, and the generated waveforms sort on average 2-3 rows.}
    \label{fig:tetris}
\end{figure*}

Each quadrant requires the rearrangement of an average of $\sim$1,500 atoms, representing a scale that has not been achieved to date. For instance, a recent example of atomic rearrangement into a 828-site array requires around 1.5~s to realize the target configuration with a single rearrangement tweezer~\cite{pichard_rearrangement_2024}. While such a timescale would slow the experimental cycle and allow some defects to be incurred due to vacuum-induced losses, algorithms that utilize many rearrangement tweezers in parallel expedite the process\cite{ebadi_quantum_2021}. Hence, we propose to use the ``Tetris algorithm''~\cite{wang_accelerating_2023}, a row-by-row sorting algorithm that capitalizes on parallelism and overcomes other timescale limitations through real-time calculation. In SI Fig.~\ref{fig:tetris}a, we show the steps of the Tetris algorithm from simulating the sorting of a randomly half-filled quadrant of our 12,000 site array into 1,445 target sites.

Averaging over $\sim$1,000 iterations of our simulation sorting one quadrant with different initial configurations, we find that the algorithm requires an average of 119 steps, with $\sim$23 atoms being moved on average in each step (see SI Fig.~\ref{fig:tetris}c). The longest distance move in the row being sorted determines the timescale of the step. In SI Fig.~\ref{fig:tetris}b we show the histogram of the longest distance move per step (with an average longest distance of $159 \ \upmu\mathrm{m}$).

\subsection{Time budget for quadrant rearrangement}

We can use our implementation of the Tetris algorithm in the previous section to give an estimate of the rearrangement time using conservative parameters based on the experimental data, simulation, and existing values in the literature. In SI Table~\ref{table:rearrangement-time} we present a summary of this time budget. We find that $\sim$137 ms would be required to rearrange quadrants in parallel, or $\sim$522 ms if the quadrants are sorted sequentially. This timescale is compatible with the operation of a typical neutral atom experiment. Since the timescales for moves assume experimental results for straight moves, which suffer from cylindrical lensing effects, the overall time could be further reduced by mitigation strategies for cylindrical lensing outlined in SI §\ref{Coherent transport}.B. In the next paragraphs we detail our rationale for each line of SI Table~\ref{table:rearrangement-time}.

\begin{table*}[t!]
\centering
\begin{tabular}{|c|c|c|c|}
\hline
\textbf{Operation} & \textbf{Time} & \textbf{Parallel sorting} & \textbf{Sequential sorting} \\
\hline
Image transfer and processing  & 10 ms & $\times1$ & $\times1$ \\
\hline
Tetris algorithm computation & 235 $\upmu$s & $\times4$ &$\times1$  \\
\hline
Waveform streaming latency  & 488 $\upmu$s & $\times4$ & $\times 1$  \\
\hline
Pick-up \& drop-off & $24.8$ ms  & $\times1$ & $\times4$\\
\hline
Tweezer splitting \& merging & $24.8$ ms & $\times1$ & $\times4$\\
\hline
Movement & $74.4$ ms & $\times1$ & $\times4$ \\
\hline
Fast-steering mirror switching & 5 ms & $\times0$ & $\times3$\\
\hline
\textbf{Total} & & ${\sim}137$ ms & ${\sim}522$ ms\\
\hline
\end{tabular}

\caption{\textbf{Time budget for quadrant rearrangement.} This conservative estimate, detailed in the SI text, is based on data presented in our work and available in the literature. We notably assume straight moves and therefore the move duration could be significantly reduced utilizing cylindrical lensing strategies detailed in SI §\ref{Coherent transport}.B. We show in the table timescales for rearranging quadrants in parallel using four sets of 2D AODs (``Parallel sorting" above) and sequentially using a fast-steering mirror to switch between quadrants (``Sequential sorting" above).  There are other options, such as partially parallelized sorting with two pairs of AODs, that would have an intermediate timescale. The pick up and drop off timescales as well as the tweezer splitting and merging timescales are calculated as following: $(62 \text{ rows} + 62 \text{ columns}) \times 200 \ \upmu$s $= 24.8$ ms. The move time is calculated as $(62 \text{ rows} + 62 \text{ columns}) \times 600 \ \upmu$s $= 74.4$ ms.}
\label{table:rearrangement-time}
\end{table*}

\medbreak
\paragraph{Image transfer and processing}\mbox{}\\
We have measured a delay of 10 ms between the end of camera exposure and the determination of the atom occupation matrix. Our implementation of the image processing algorithm presented in the main text is written in Python, and we rely on the fast CoaXPress connection with the Hamamatsu qCMOS camera. This delay time will be the same regardless of the sorting implementation.

\medbreak
\paragraph{Tetris algorithm computation}\mbox{}\\
We have written an implementation of the Tetris algorithm for our array using C++. Owing to its relative simplicity, we can calculate and determine which moves to execute for rearrangement of one quadrant in about 235 $\upmu$s. In the case of sequential sorting, this calculation could be done during the fast-steering mirror switching, so we only include the calculation of the first quadrant sorting in the table.

\medbreak
\paragraph{Waveform streaming latency}\mbox{}\\
We turn to estimating the latency associated with waveform generation and streaming to the AODs. Results in the literature vary depending on the hardware and the implementation. Examples include synthesis from a computer-controlled arbitrary waveform generation card~\cite{endres_atom-by-atom_2016} or generation from a field-programmable gate array equipped with a digital-to-analog converter~\cite{young_half-minute-scale_2020,wang_accelerating_2023}. For simplicity, one may store in memory all the waveforms associated with moving a single trap~\cite{madjarov_entangling_2021}, and during the experiment sequence sum the appropriate waveforms. For the atom array introduced in this work, we estimate that we would need to store 2.53 million waveforms. Assuming a realistic sampling rate and average waveform duration, this represents over 250 billion samples, or more than 1 terabyte of data (with single-precision floating-point numbers). One would need to sum these waveforms on the fly in addition to storing this immense amount of data, which is essentially prohibitive to the practicality of this approach.

Our approach provides an alternative to storing waveforms. We write and benchmark a real-time waveform generation algorithm from scratch on a consumer graphics processing unit (GPU, NVIDIA GeForce RTX 3090), leveraging the device's high calculation throughput using the C++ CUDA platform.
The waveform samples are grouped in ``chunks" for streaming, while the subsequent chunk is generated. The latency is defined as the time that it takes to calculate the first chunk before the waveform generation is simultaneous to streaming. The chunk size of 524k samples is chosen, which on average covers the waveforms for sorting 2-3 rows of a quadrant using 64 moving tweezers (slightly more than the number of rows in a quadrant). This amount of data represents about 2.6 ms of waveform streaming, and the time it takes to generate each chunk is on average ${\sim}488 \ \upmu\mathrm{s}$ (the full histogram is shown in SI Fig.~\ref{fig:tetris}d).

Since the generation time is shorter than the waveform durations, one can stream the first waveform chunk to the arbitrary waveform generator, compute the second chunk and stream it before the first one ends, and so on. This enables continuous streaming with latency given by the generation time of the first chunk, on average $488\ \upmu$s.

In the time budget in SI Table~\ref{table:rearrangement-time}, for the case of sequential sorting, we count this this latency time once, given that after the moves are generated for the first quadrant, this calculation can be done during the fast-steering mirror switching time of 5 ms. For the case of parallel sorting, we count this latency four times as a conservative estimate, although this could be also implemented in parallel by utilizing additional workstations or GPUs.

\begin{figure*}[hbt!]
	\centering
	\includegraphics[width=90mm]{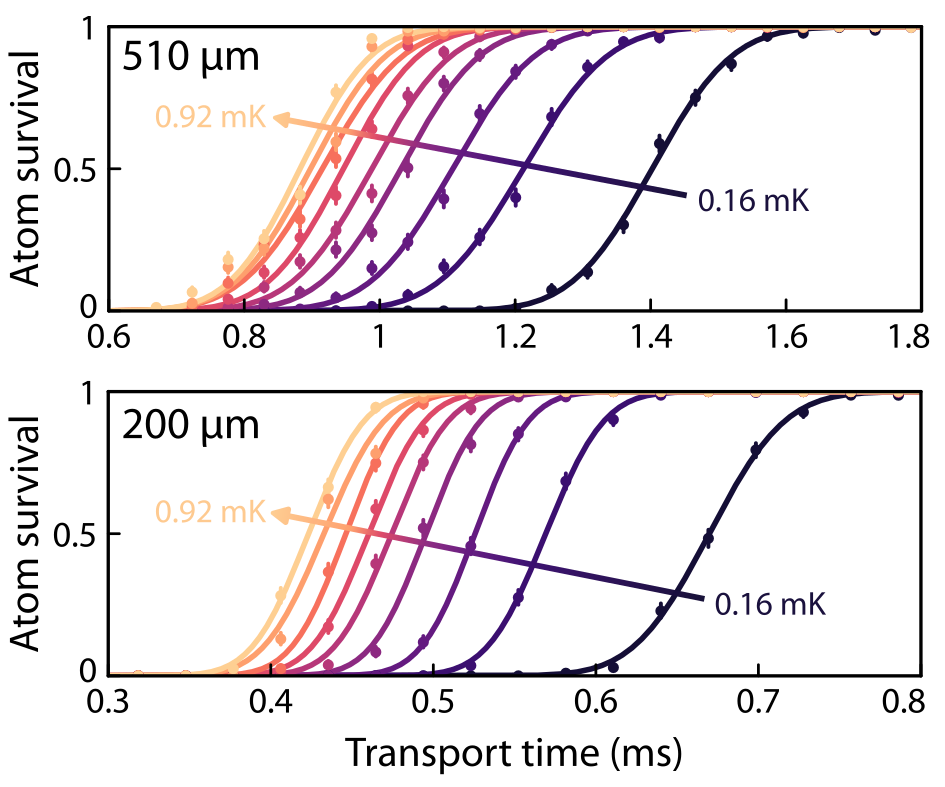}
	\caption{\textbf{Survival during transport: straight moves.} Experimentally measured atom survival for different trap depths and two distances of 510 $\upmu$m and 200 $\upmu$m. The signal for each depth is fitted to an error function. Atoms are transported using the constant jerk trajectory with a straight, single-axis move as opposed to the diagonal moves used in the coherent transport study in the main text (Fig.~5a). Note that these moves could be executed considerably faster by mitigating cylindrical lensing, as discussed in SI §\ref{Coherent transport}.A.}
    \label{fig:transport-time}
\end{figure*}

\medbreak
\paragraph{Pick-up \& drop-off , splitting \& merging}\mbox{}\\
The times needed to pick up and drop off an atom from a static to a transport tweezer, and to split and merge the static and transport tweezers are not dependent on the distance of the rearrangement move being performed. In Fig.~6d we present data showing atom survival for repeated (coherent transfer): we notably find that a single one-way transfer can be done in 200 $\upmu$s with a survival probability of more than 98\%. Moreover, higher AOD and SLM trap depths could be used to further speed up this operation---since for rearrangement we are not concerned with qubit coherence. Hence, we pick this timescale as a reasonable assumption for rearrangement. As is typical for smaller-scale experiments, one could perform a second, much shorter rearrangement round to further increase the success probability. The Tetris algorithm first implements row-based rearrangement of atoms, with one pick-up, split, merge, and drop-off operation for every row, and then compresses atoms in columns with one with one pick-up, split, merge, and drop-off operation per column: hence, we multiply by (62 rows + 62 columns) in the Table.

\medbreak
\paragraph{Movement}\mbox{}\\
Regarding transport itself, it would be more difficult to extrapolate from previous results since there are important trade-offs associated with long-distance moves, as discussed in the main text and in SI~§III.A. Hence, we instead experimentally measure the required transport times for two different distances and various trap depths for vertical moves (SI Fig.~\ref{fig:transport-time}). We then select the average transport time, 600 $\upmu$s, as the transport time required to obtain atom survival $\gtrsim 0.999$ using a tweezer with depth $0.28$ mK for a distance of 200 $\upmu$m, which is slightly further than the average longest distance per set of parallel moves found by simulating the Tetris algorithm. Note that, as discussed in the main text and the Methods, this timescale could be sped up by methods for mitigating cylindrical lensing (SI §\ref{Coherent transport}.B), with a deeper trap depth, or with machine learning, but we choose to base our estimate off of this value.

\medbreak
\paragraph{Fast-steering mirror switching}\mbox{}\\
As mentioned above, one can obtain a Thorlabs fast-steering mirror with a switching time of 5 ms. We include this time in our estimation of the sequential sorting sequence which requires the field of view of the AODs to switch between quadrants.

\section{Coherent transport between zones and hyperfine control}
\label{Coherent transport}
Coherent transport of atoms~\cite{beugnon_two-dimensional_2007} during the quantum processor operation is a key tool that enables all-to-all connectivity with low overhead in recent experiments that implement a universal quantum processor with an atom array~\cite{bluvstein_quantum_2022,bluvstein_logical_2024}. While there are some challenges associated with scaling up the size of the array, we demonstrate in our work that high-fidelity moves can be realized even for a distance of 610 $\upmu$m as shown in Fig.~5d. Importantly, this distance is greater than the largest distance needed to move atoms between storage and interaction or readout zones in our proposed layout. In this work, we found that we are able to move much more quickly with a diagonal motion when compared with a straight motion (Fig.~5a). We attribute this to the impact of cylindrical lensing, which we examine more closely in the next section, particularly highlighting the way this effect becomes more detrimental when scaling to AODs with a larger field of view.
In this work, we have also examined coherent trap-transfer between static SLM traps and dynamic AOD traps at the scale of 195 tweezers across 504 $\times$ 468 $\upmu$m (Fig. 6). We go on to combine this with a the long-distance coherent moves that we demonstrate, by characterizing the fidelity of trap-transfer for 47 tweezers, followed by a $375 \mu$m move between the upper and lower half of a large-scale static array of tweezers (Ext. Data Fig.~10e,f). Further development of the fidelity and speed of this operation is expected with the use of machine learning techniques demonstrated in Fig.~6b and c.

\subsection{Critical speed from cylindrical lensing}
When moving the atoms longer distances, cylindrical lensing becomes an increasingly pronounced problem, determining a critical speed that is impossible to exceed. As the tweezer is moved using a changing radio-frequency wave in the AOD crystal, the focus of the tweezer is offset from the atom plane by the distance $z_s = \frac{f}{V} \dot{x}_0(t)$ where $f$ is the effective focal length of the objective (including any magnification after the AOD), $V$ is the velocity of the acoustic wave in the AOD, and $\dot{x}_0(t)$ is the tweezer velocity. This results in a modified tweezer potential $U \propto -\tilde{U}_x \tilde{U}_y$ that can be described, when moving in the $x$ direction as 
$$ \tilde{U}_x = \frac{1}{\left(1+(\frac{z-z_S}{z_R})^2 \right)^{1/2}} \exp{\left(-\frac{2x^2}{w_0^2 \left(1+(\frac{z-z_S}{z_R})^2\right)} \right)},$$

$$ \tilde{U}_y = \frac{1}{\left(1+(\frac{z}{z_R})^2 \right)^{1/2}} \exp{\left(-\frac{2y^2}{w_0^2 \left(1+(\frac{z}{z_R})^2 \right)} \right)}.$$

The ratio of $\abs{z_s}$ to $z_R$ (the tweezer Rayleigh range) determines two regimes: when $\abs{z_S} < 2z_R$, the tweezer is still well defined (i.e., there is a single local potential minimum, albeit broadened by cylindrical lensing) and the trap depth is reduced by $1+\frac{1}{4}\left(\frac{z_s}{z_R} \right)^2$. When the tweezer is moved too quickly in one direction and $\abs{z_s}$ exceeds $2z_R$, the trap splits such that there are two potential minima, causing the most deleterious affects. This motivates us to introduce a \emph{characteristic shift velocity}, $v_s$, such that $\frac{z_s}{2z_R} = \frac{\dot{x}_0}{v_s}$. Notably, $v_s$ constitutes a velocity that is difficult to exceed during atom transport without atomic loss, and is given by: $v_s = 2\pi V w_0^2/f\lambda$. By assuming Airy diffraction through the objective, we obtain a simple formula:

\[v_s \approx 5.3 \frac{w_0}{T_a},\]

where $T_a$ is the AOD access time, i.e. the propagation time of the acoustic wave through the input beam~\cite{goutzoulis_design_2021}. (Note that the exact factor depends on the optical configuration; in our case, we obtain results consistent with a factor $7-8$ instead of 5.3). This formula highlights an important trade-off: $T_a$ factors in the time-bandwidth product ($T_a \, \Delta f)$, a figure of merit for AODs that is roughly equal to the number of resolvable spots. Thus, as one tries to scale up the addressable field of view of AODs, $v_s$ decreases.

Using a diagonal trajectory allows this splitting to be eliminated, minimizing the impact of lensing, though there remains a penalty due to the off-axis focus shift. 

\subsection{Cylindrical lensing mitigation for fast coherent transport}

In the next paragraphs, we highlight several directions to mitigate the deleterious effects due to cylindrical lensing. Implementing such techniques would enable faster high-fidelity transport, and allow for more effective dynamical decoupling by maintaining a consistent trap depth during motion.

\medbreak

\paragraph{Diagonal motion}\mbox{}\\
To mitigate the effects of cylindrical lensing in the data that we present in the main text, we have used diagonal moves (by which we mean that tweezers move with the same rate along $x$ and $y$; $\dot{f_x} = \dot{f_y}$), which turns cylindrical lensing into spherical lensing. While there is still a penalty associated with the focal shift of the tweezer, it is much reduced compared with the straight move situation. We have leveraged this strategy to realize faster high-fidelity moves, as presented in Fig.~5 in the main text, and in Ext. Data Fig.~10e,f.

\begin{figure*}[hbt!]
	\centering
	\includegraphics[width=80mm]{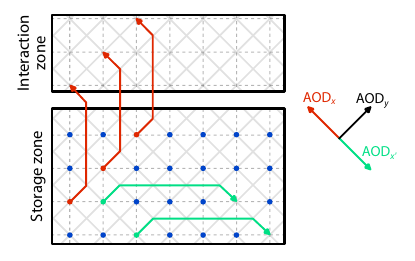}
	\caption{\textbf{Diagonal-move based zone architecture.} Schematic of a zone layout that is compatible with diagonal motion with AODs. Each arrow associated with a distinct AOD indicates the diffraction axis in the direction of increasing acoustic frequency. To realize the vertical (red) set of moves, one would use $\mathrm{AOD}_x$ and $\mathrm{AOD}_y$, and for the horizontal (green) set, $\mathrm{AOD}_{x'}$ and $\mathrm{AOD}_y$}
    \label{fig:diag-zones}
\end{figure*}

In SI Fig.~\ref{fig:diag-zones} we describe how a zone-based neutral atom quantum processor can be engineered with these diagonal moves. Essentially, there is no drawback associated with configuring transport tweezers in a diagonal way, aside from the need to conceive blocks of atoms --- e.g., a block encoding a logical qubit --- in a diagonal way. We have already demonstrated a preliminary such diagonal orientation of static tweezers in Ext. Data Fig.~10e which we use for characterization of the fidelity of combining coherent pick-up with a $375\mu$m move.

Importantly, this setup, allowing arbitrary moves for dynamic tweezers, requires three AODs. For instance, one may use two orthogonal AODs, $\mathrm{AOD}_x$ and $\mathrm{AOD}_y$, to move in the vertical direction, such that along this direction $\dot{f_x} = \dot{f_y}$. If this configuration was used to move tweezers along the horizontal direction, then $\dot{f_x} = -\dot{f_y}$, which would maximize detrimental effects from cylindrical lensing. To remove lensing effects for movements in the vertical-horizontal grid, another AOD is required, with a parallel opposite deflection axis with respect to one of the two other AODs.

\medbreak

\paragraph{Cylindrical lensing suppression with counter-propagating wave AODs}\mbox{}\\
An alternative solution that has been proposed to strongly suppress acousto-optic lensing is to use two acousto-optic deflectors (for each dimension) stacked so that the acoustic wave in the second AOD is identical but counter-propagating with respect to the first AOD~\cite{friedman_acousto-optic_2000}. 
By diffracting the tweezer light as the respective $(+1)$ and $(-1)$ orders of each AOD, the diffraction angles add up but the lensing effects cancel to first order. Compared with diagonal motion, this presents the added benefit of removing any lensing (including spherical). While there does not seem to be any \emph{a priori} obstacle to implementing this scheme with a large-FOV tweezer array, the in-situ performance remains to be characterized.

\medbreak

\paragraph{Acousto-optical deflectors with faster access times}\mbox{}\\
As discussed in the previous section, the characteristic shift velocity for a given AOD and optical configuration is given by $v_s/w_0 \sim 5.3/T_a$, where $w_0$ is the tweezer waist and $T_a$ the \emph{access time} (or \emph{time aperture}) of the AOD. $T_a$ factors in a crucial figure of merit of wideband AODs, the time-bandwidth product $T_a \, \Delta f$, which describes up to a unit-scale factor the number of resolvable spots~\cite{goutzoulis_design_2021}. Therefore the parameter space to decrease $T_a$ while preserving the number of resolvable spots and the scan angle is highly constrained. Since all commercially available AODs at near-infrared wavelengths currently use the shear mode of tellurium dioxide (which possesses an unusually low acoustic velocity), for which further improvement in the figures of merit mentioned above seem unrealistic, it appears that further research in acousto-optic materials would be required to fabricate wide-band AODs at near-infrared wavelengths with low access time.

\subsection{Global single-qubit gates with a Raman amplitude-modulation setup}

We demonstrate in the main text high-fidelity single-qubit gates using a global microwave drive. However, it is desirable to achieve much faster Rabi frequency, which requires optical addressing through Raman transitions~\cite{yavuz_fast_2006,jones_fast_2007}. This technique is also amenable to single-site addressing, which we discuss in the next subsection. While a well-established technique, scaling Raman single-qubit gates to the large atom array introduced in the main text may present challenges due to potential limitations in the available laser power. For this reason, we propose to use the amplitude-modulation setup developed in Ref.~\cite{levine_dispersive_2022} This scheme is particularly power-efficient and does not require low laser phase noise. The effective Rabi frequency is given by:

$$\Omega_{\mathrm{eff}} = \eta_1 \frac{|\Omega|^2}{2\Delta},$$

where $\Omega$ is the single photon Rabi frequency from the ground state to the intermediate state ($6 P_{1/2}$ in our case), $\Delta$ is the single-photon detuning, and $\eta_1$ is the amplitude-modulation efficiency. $\eta_1$ can comfortably reach its maximum value of ${\sim}0.58$ using a resonant electro-optic modulator and a volumetric chirped Bragg grating.

This setup was used to perform Raman sideband spectroscopy on the atom array, presented in Ext. Data Fig.~8. Using a titanium:sapphire ring laser at 895 nm, we obtain 1 W of fiber-coupled amplitude-modulated laser light, with a single-photon detuning $\Delta = 2\pi \times 345$ GHz. With these parameters, we estimate that an effective Rabi frequency of $2\pi \times 2$ MHz is achievable, with beam-shaping chosen such that inhomogeneity across the $900 \ \upmu$m tweezer array is low enough to achieve a single-qubit gate fidelity of more than 0.9999 with a composite pulse such as SCROFULOUS~\cite{cummins_tackling_2003} or BB1~\cite{wimperis_broadband_1994}. Alternatively, one could address a specific zone of the array using precise beam shaping techniques with an SLM~\cite{bowman_high-fidelity_2017,ebadi_quantum_2021,swan_high-fidelity_2025}.

\subsection{Local single-qubit gate addressing}

While the capacity to transport atoms is a crucial feature of atom array quantum computation, it is also desirable to be able to apply single-qubit gates to single atoms or patches of atoms, for applications to implementing local gates or locally-tailored dynamical decoupling for transported atoms. In order to utilize Raman addressing through an objective and perpendicular to a quantization axis, solutions to challenges associated with the quantization axis position with respect to the objective axis have been proposed in previous work~\cite{bluvstein_logical_2024}. Beyond this, we discuss here additional potential adaptations for implementation of local addressing at the scale of the atom array.

We consider, for instance, local addressing with a pair of crossed AODs. This technique, in addition to being relatively experimentally simple, allows for parallel addressing, provided the addressing pattern obeys the product structure imposed by the AODs. While we mention previously that no commercially available AODs can cover the full atom array, the addressing beams need not be diffraction-limited. Hence, one can further demagnify AOD-diffracted beams in order to cover a larger field of view, up to the resolution limit of the AOD. The previously mentioned AOD model has a resolution of ${\sim}500$ resolvable spots, large enough to create individual addressing spots for each site.

Alternatively, a novel optical scheme proposed and demonstrated in Ref.~\cite{zhang_scaled_2024} utilizes a combination of SLM and DMD technology to generate an array of over 10,000 spots that can be quickly switched in arbitrary combinations. Such a scheme, already demonstrated optically at large scale, would allow for \emph{fully parallel} local qubit rotations across the array. Finally, there is an ongoing research effort to scale up nanophotonic chips based on an array of Mach-Zehnder interferometers, enabling fast and arbitrary addressing of many atoms~\cite{menssen_scalable_2023}. In the future, these devices could enable ultra-fast switching and high-fidelity arbitrary addressing of large atom arrays.

\section{Rydberg gates}
\label{Rydberg Gates}
\begin{figure*}[hbt!]
	\centering
	\includegraphics[width=50mm]{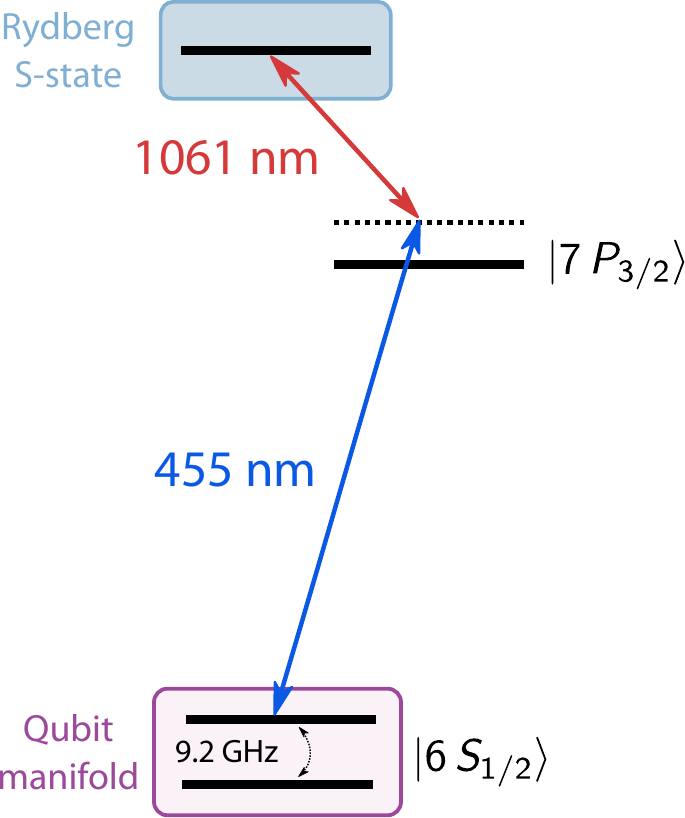}
	\caption{\textbf{Rydberg excitation pathway.} We picture the two-photon excitation pathway for Cs through the $7P_{3/2}$ state, consisting of a 455 nm and 1061 nm photon. We have available 120 W of power at 1061 nm and 2.5 W of power at 455 nm.}
    \label{fig:rydberg}
 \end{figure*}
 
To implement two-qubit gates in the hyperfine-based qubits, we plan to utilize Rydberg interactions to engineer $\mathcal{C \! Z}$ gates\cite{levine_parallel_2019,isenhower_demonstration_2010,wilk_entanglement_2010,jaksch_fast_2000}.
To execute two-qubit gates in parallel during quantum processor operations, we design an interaction zone as depicted in SI Fig.~\ref{fig:zone-architecture}.

We plan to utilize the two-photon excitation pathway pictured in the level scheme diagram in SI Fig.~\ref{fig:rydberg} through the $7P_{3/2}$ intermediate state. In order to achieve uniform two-photon Rabi frequency across the zone and to minimize the cross-talk between the Rydberg beam and qubits in the storage zone, we plan to employ spatial light modulators (SLMs) to shape the Rydberg beams into a flat-top profile~\cite{bowman_high-fidelity_2017,ebadi_quantum_2021,swan_high-fidelity_2024}, instead of Gaussian-shaped beams. We plan to use beam dimensions of approximately 100 $\upmu$m in the atom-plane and 30 $\upmu$m along the axis perpendicular to the plane.
Our setup includes a laser system with an output power of 120 W at 1061 nm sourced from a fiber-amplified laser, and an output power of 2.5 W at 455 nm. To minimize scattering from the intermediate state ($7P_{3/2}$) during the two-photon Rydberg excitation, we will opt for a significant detuning from the state.

We take into account laser power losses along the optical path from components such as acousto-optic modulators (AOMs) and SLMs~\cite{swan_high-fidelity_2024} assuming that ${\sim}20-25\%$ of the output from the lasers reaches the atoms (different for the two paths taking into account wavelength-dependent diffraction efficiencies). For an intermediate state detuning of $-2$ GHz, we estimate achieving an effective two-photon Rabi frequency of approximately $6.9$ MHz using the $60S_{1/2}$ Rydberg state. 

For the implementation of the two-qubit gate using Rydberg blockade, we plan to position two interacting atoms 2.5 $\upmu$m apart using two tweezers created by lasers of different wavelengths to avoid interference (see Ext. Data Fig.~3c). Each pair will be spaced by $11.4$ $\upmu$m from adjacent pairs. We note that at this large spacing of tweezers from the same laser source, out-of-plane interference is minimized (Ext. Data Fig.~3a). This configuration is anticipated to yield a blockade shift of around 300 MHz for the paired atoms and a residual interaction strength of ${\sim}0.03$ MHz on neighboring sites, comparable to the residual interaction strengths in other contemporary experiments with state of the art two-qubit gate fidelities~\cite{evered_high-fidelity_2023,tsai_benchmarking_2025}. With these settings, we expect that state-of-the-art results for two-qubit gate fidelities can be reproduced at the scale of ${\sim}500$ two-qubit gate sites, in the ${\sim}110$ $\upmu$m by ${\sim}900$ $\upmu$m zone.

\section{Beyond array size limitations}
\label{Beyond Array Size Limitation}
After mitigating the objective heating concerns that limit the amount of power that we can use for tweezers, we envision utilizing the full power out of the second fiber amplifier at 1055 nm to create more traps. In particular, we plan to interleave traps of two wavelengths, which would allow for the creation of more traps at tighter spacing without concerns of SLM phase interference, as seen in Ext. Data Fig.~3. Such tighter spacing allows for efficient usage of the objective field of view, in addition to maximizing SLM diffraction efficiency. We also plan to replace the D-mirror 
0\textsuperscript{th}-order SLM filtering on the 1061 nm tweezer path with a mirror with a hole would allow interleaving of tweezers across the whole array.
To combine these paths along with rearrangmement tweezers as mentioned previously, we also plan to replace the polarizing beam splitting cube used for combination of the pathways with a dichroic mirror. One could envision in the long term stacking multiple more dichroic mirrors in order to scale the number of tweezers even further by adding more arrays of different wavelengths in succession. Given that fiber amplifiers can be obtained across a wide range of wavelengths (around 1020-1080 nm) with current technology, one could imagine combining over 10 pathways separated by around 5 nm each. 

\section{Summary}
\label{Summary}
We conclude here that the comprehensive nature of the planned implementation laid out above provides promising evidence for straightforward progress in the near term towards universal quantum computing at the scale of $\sim$6,000 atoms. This corroborates the discussion in the main text of ``A tweezer array with 6100 highly coherent atomic qubits", with exciting consequences for the scalability of the platform and usability of neutral atom processors for quantum error correction of many logical qubits.


\bibliography{e2-references,additional-ref}